\documentclass[journal]{IEEEtran}

\usepackage{cite}
\usepackage{graphicx}
\usepackage[lowtilde]{url}
\usepackage[caption=false,font=footnotesize]{subfig}
\usepackage{grffile}
\usepackage{amsmath}
\usepackage{multirow}
\usepackage{enumitem}
\usepackage{amssymb}
\usepackage[ruled,norelsize]{algorithm2e}
\usepackage[euler]{textgreek}
\usepackage{booktabs} 
\usepackage{array}

\makeatletter
\long\def\@makecaption#1#2{\ifx\@captype\@IEEEtablestring%
\footnotesize\begin{center}{\normalfont\footnotesize #1}\\
{\normalfont\footnotesize\scshape #2}\end{center}%
\@IEEEtablecaptionsepspace
\else
\@IEEEfigurecaptionsepspace
\setbox\@tempboxa\hbox{\normalfont\footnotesize {#1.}~~ #2}%
\ifdim \wd\@tempboxa >\hsize%
\setbox\@tempboxa\hbox{\normalfont\footnotesize {#1.}~~ }%
\parbox[t]{\hsize}{\normalfont\footnotesize \noindent\unhbox\@tempboxa#2}%
\else
\hbox to\hsize{\normalfont\footnotesize\hfil\box\@tempboxa\hfil}\fi\fi}
\makeatother

\newcommand{\RN}[1]{%
  \textup{\uppercase\expandafter{\romannumeral#1}}%
}

\makeatletter
\newcommand{\removelatexerror}{\let\@latex@error\@gobble}
\makeatother

\newcolumntype{C}[1]{>{\centering\let\newline\\\arraybackslash\hspace{0pt}}m{#1}}

\begin{document}

\title{Modeling Generalized Rate-Distortion Functions}

\author{Zhengfang~Duanmu,~\IEEEmembership{Student Member,~IEEE,}\\
        Wentao~Liu,~\IEEEmembership{Student Member,~IEEE,}
        and~Zhou~Wang,~\IEEEmembership{Fellow,~IEEE}
\thanks{ The authors are with the Department of Electrical and Computer Engineering, University of Waterloo, Waterloo, ON N2L 3G1, Canada (e-mail: \{zduanmu, w238liu, zhou.wang\}@uwaterloo.ca).}
}

\markboth{}%
{Shell \MakeLowercase{\textit{et al.}}: Bare Demo of IEEEtran.cls for Journals}

\maketitle

\begin{abstract}
Many multimedia applications require precise understanding of the rate-distortion characteristics measured by the function relating visual quality to media attributes, for which we term it the generalized rate-distortion (GRD) function. In this study, we explore the GRD behavior of compressed digital videos in a three-dimensional space of bitrate, resolution, and viewing device/condition. Our analysis on a large-scale video dataset reveals that empirical parametric models are systematically biased while exhaustive search methods require excessive computation time to depict the GRD surfaces. By exploiting the properties that all GRD functions share, we develop an Robust Axial-Monotonic Clough-Tocher (RAMCT) interpolation method to model the GRD function. This model allows us to accurately reconstruct the complete GRD function of a source video content from a moderate number of measurements. To further reduce the computational cost, we present a novel sampling scheme based on a probabilistic model and an information measure. The proposed sampling method constructs a sequence of quality queries by minimizing the overall informativeness in the remaining samples. Experimental results show that the proposed algorithm significantly outperforms state-of-the-art approaches in accuracy and efficiency. Finally, we demonstrate the usage of the proposed model in three applications: rate-distortion curve prediction, per-title encoding profile generation, and video encoder comparison.
\end{abstract}


\IEEEpeerreviewmaketitle

\section{Introduction}\label{sec:introduction}

\IEEEPARstart{R}{ate-distortion} theory provides the theoretical foundations for lossy data compression and are widely employed in image and video compression schemes. Many multimedia applications require precise measurements of rate-distortion functions to characterize source signal and maximize user Quality-of-Experience (QoE). Examples of applications that explicitly use rate-distortion measurements are codec evaluation~\cite{grois2013performance}, rate-distortion optimization~\cite{wang2012ssim}, video quality assessment (VQA)~\cite{ou2014q}, encoding representation recommendation~\cite{zhang2013qoe,toni2015optimal,de2016complexity,chen2016subjective}, and QoE optimization of streaming videos~\cite{wang2015objective,chen2017encoding}.

\begin{figure*}
  \centering
  \includegraphics[width=0.9\textwidth]{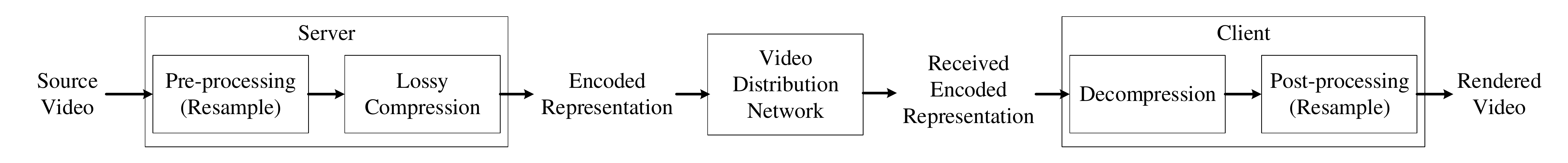}
  \caption{Flow diagram of video delivery chain.}\label{fig:pipline}
\end{figure*}

Digital videos usually undergo a variety of transforms and processes in the content delivery chain, as shown in Fig.~\ref{fig:pipline}. To address the growing heterogeneity of display devices, contents, and access network capacity, source videos are encoded into different bitrates, spatial resolutions, frame rates, and bit depths before transmitted to the client. In an adaptive streaming video distribution environment~\cite{ISO2012Dash}, based on the bandwidth, buffering, and computation constraints, client devices adaptively select a proper video representation on a per-time segment basis to download and render. Each process influences the visual quality of a video in a different way, which can be jointly characterized by a generalized rate-distortion (GRD) function. In general, this attribute-distortion mapping comprises several complex factors, such as source content, operation mode/type of encoder, rendering system, and human visual system (HVS) characteristics. In this work, we assume the GRD surface is a function $f: \mathbb{R}^2 \to \mathbb{R}^J$, where the input of the function is the video representation consisting of bitrate and spatial resolution, the output of the function is the perceptual video quality at different viewing condition, and $J$ represents the number of viewing devices, respectively. Furthermore, the GRD function is content- and encoder-dependent.

Despite the tremendous growth in computational multimedia over the last few decades, estimating a GRD function is difficult, expensive, and time-consuming. Specifically, probing the quality of a single sample in the GRD space involves sophisticated video encoding and quality assessment, both of which are expensive processes. For example, the recently announced highly competitive AV1 video encoder~\cite{aom2018AV1} and video quality assessment model VMAF~\cite{li2016VMAF} could be over 1000 times and 10 times slower than real-time for full high-definition (1920$\times$1080) video content. Given the massive volume of multimedia data on the Internet, the real challenge is to produce an accurate estimate of the GRD function with a minimal number of samples.

We aim to develop a GRD function estimation framework with three desirable properties:
\begin{itemize}
  \item Accuracy: It produces asymptotically unbiased estimation of GRD function, independent of the source video complexity and the encoder mechanism.

  \item Speed: It requires a minimal number of samples to reconstruct a full GRD function.

  \item Mathematical soundness: The GRD model has to be mathematically well-behaved, making it readily applicable to a variety of computational multimedia applications.
\end{itemize}
\begin{figure*}[t]
    \centering
    \captionsetup{justification=centering}
    \subfloat[]{\includegraphics[width=0.33\textwidth]{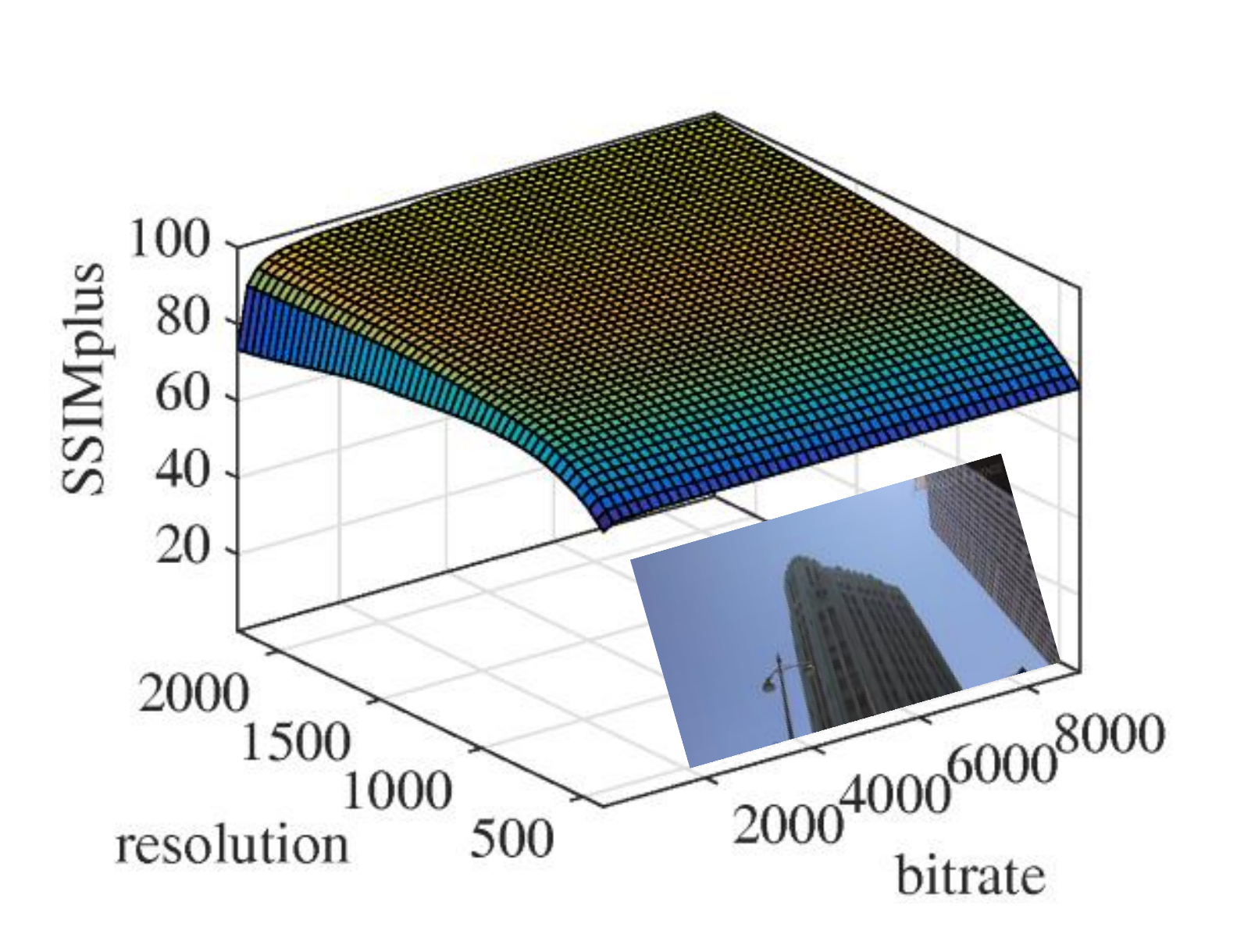}}\hskip.2em
    \subfloat[]{\includegraphics[width=0.33\textwidth]{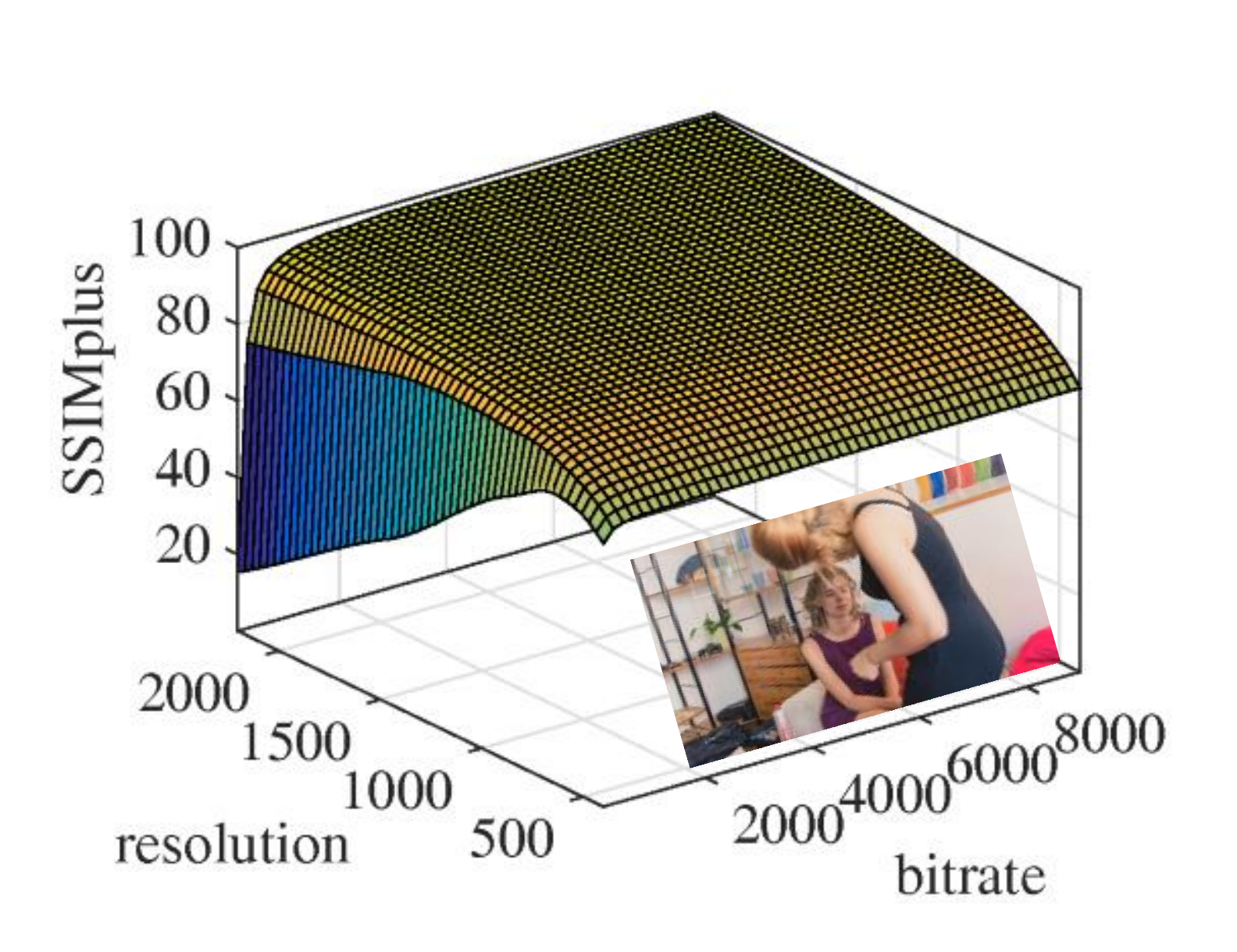}}\hskip.2em
    \subfloat[]{\includegraphics[width=0.33\textwidth]{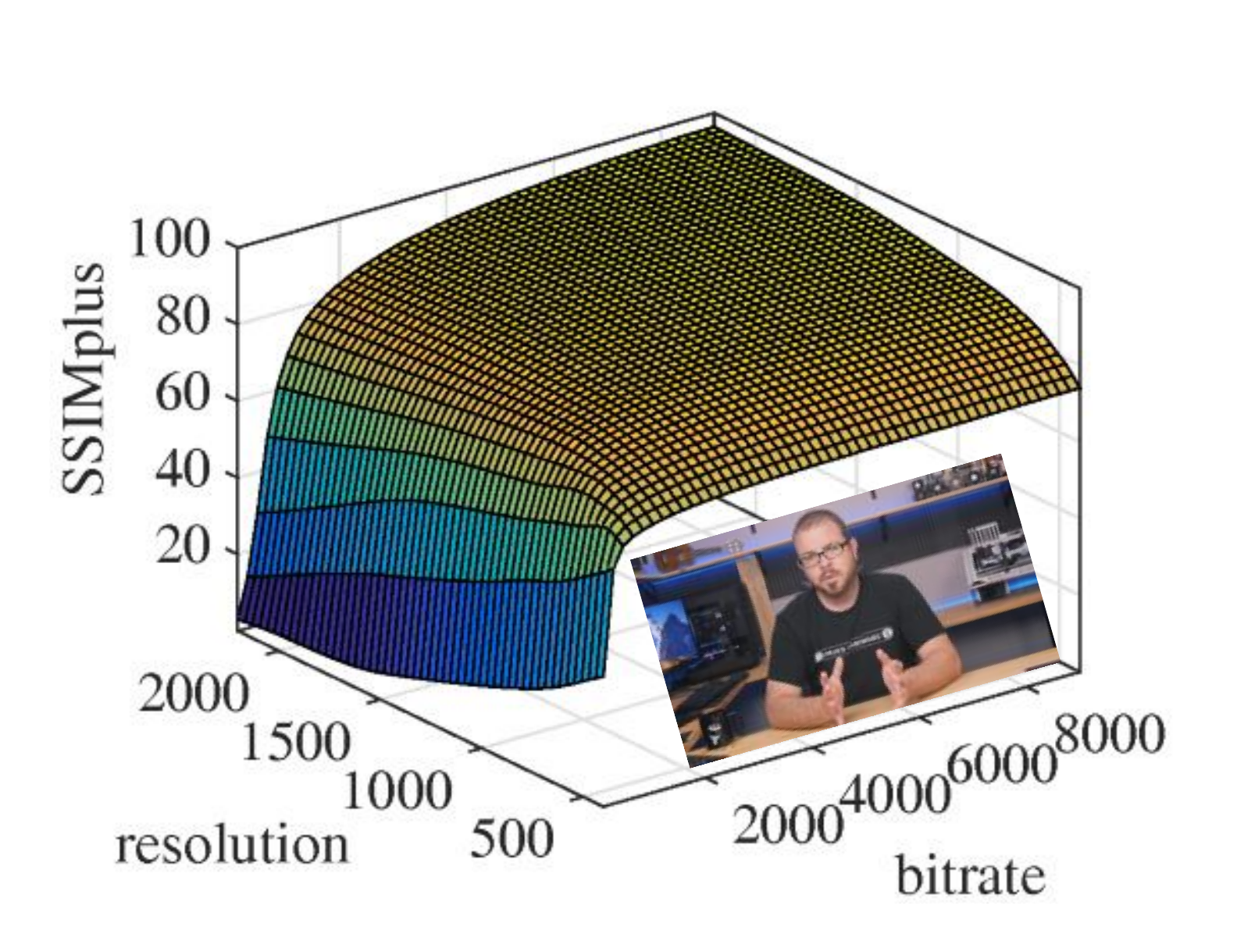}}
    \caption{Samples of generalized rate-distortion surfaces for different video content.}\label{fig:samples}
\end{figure*}
To achieve \textit{accuracy}, we analyze the properties that all GRD functions share, based on which we formulate the GRD function approximation problem as a quadratic programming problem. The solution of the optimization problem provides an optimal interpolation model lying in the theoretical GRD function space. To achieve \textit{speed}, we propose an efficient sampling algorithm that constructs a set of queries to maximize the expected information gain. The sampling scheme results in a unique sampling sequence invariant to source content, enabling parallel encoding and quality assessment processes. To achieve \textit{mathematical soundness}, the GRD model is inherited from the Clough-Toucher (CT) interpolation method, and the function is differentiable everywhere on the domain of interests. Extensive experiments demonstrate that the resulting GRD function estimation framework achieves consistent improvement in speed and accuracy compared to the existing methods. The superiority and usefulness of the proposed algorithm are also evident by three applications.

\section{Related work}\label{sec:literature}
Although rate-distortion theory has been successfully employed in many multimedia applications, the research in the GRD surface modeling has only become a scientific field of study in the past decade. Existing methods can be roughly categorized based on their assumptions about the shape of a GRD function. The first model class only makes weak assumptions about the properties of the GRD functions. For example, \cite{de2016complexity} assumes the continuity of GRD functions and apply linear interpolation to estimate the response function after densely sampling the video representation space. However, the exhaustive search process is computationally expensive, not to mention the number of samples required increases exponentially with respect to the dimension of input space.

By contrast, the second class of models make strong \textit{a priori} assumptions about the form of the GRD function to alleviate the need of excessive training samples. For example,~\cite{ou2014q} assumes the video quality exhibits an exponential relationship with respect to the quantization step, spatial resolution, and frame rate. Alternatively, Toni~\textit{et al.}~\cite{toni2015optimal,kreuzberger2016comparative} derived a reciprocal function to model the GRD function. Similarly,~\cite{chen2016subjective} modeled the rate-quality curve at each spatial resolution with a logarithmic function. A significant limitation of these models is that domains of the analytic functional forms are restricted only to the bitrate dimension and several discrete resolutions, lacking the flexibility to incorporate other dimensions such as frame rate and bit depth, and the capability to predict the GRD behaviors at novel resolutions.

In addition to the specific limitations the two kinds of models may respectively have, they suffer from the same problem that the training samples in the GRD space are either manually picked or randomly selected, neglecting the difference in the informativeness of samples. While many recent works acknowledge the importance of GRD function~\cite{zhang2013qoe,toni2015optimal,de2016complexity,chen2016subjective}, a careful analysis and modeling of the response has yet to be done. We wish to address this void. In doing so, we seek a good compromise between 1) global and rigid models depending on random training samples and 2) local and indefinite models requiring exhaustive search in the video representation space.

\section{Modeling Generalized Rate-Distortion Functions}\label{sec:model}
We begin by stating our assumptions. Our first assumption is that the GRD function is smooth. In theory, the Shannon lower bound, the infimum of the required bitrate to achieve a certain quality, is guaranteed to be continuous with respect to the target distortion~\cite{berger1975rate}. On the other hand, successive change in the spatial resolution would gradually deviate the frequency component and entropy of the source signal, resulting in smooth transition in the perceived quality. In practice, many subjective experiments have empirically shown the smoothness of GRD functions~\cite{ou2014q,zhai2008cross}. Furthermore, it is beneficial to impose $C^1$ continuity on GRD functions for better mathematical properties. For instance, the GRD surface is desired to be differentiable in many multimedia applications~\cite{toni2015optimal,chen2017encoding}.

Our second assumption is that the GRD function is axial-monotonic\footnote{In this work, we use rate-distortion function and rate-quality function interchangeably. Without loss of generality, we assume the function $f$ to be axial monotonically increasing. If $f$ is axial monotonically decreasing, we replace the given response with the function $f_{max}-f$, where $f_{max}$ is the maximum value of quality.}. According to the rate-distortion theory~\cite{berger1975rate}, video quality increases monotonically with respect to the amount of resources it takes in the lossy compression. However, such monotonicity constraint may not apply to the spatial resolution. It has been demonstrated that encoding at high spatial resolution may even result in lower video quality than encoding at low spatial resolution under the same bitrate combined with upsampling and interpolation~\cite{de2016complexity}. To be specific, encoding at high resolution with insufficient bitrate would produce artifacts such as blocking, ringing, and contouring, whereas encoding at low resolution with upsampling and interpolation would introduce blurring. The resulting distortions are further amplified or alleviated by the characteristics of the viewing device and viewing conditions, which interplay with HVS features such as the contrast sensitivity function~\cite{robson1966spatial}. A few sample GRD surfaces with their corresponding source videos are illustrated in Fig.~\ref{fig:samples}.

\begin{figure}
  \centering
  \includegraphics[width=0.40\textwidth]{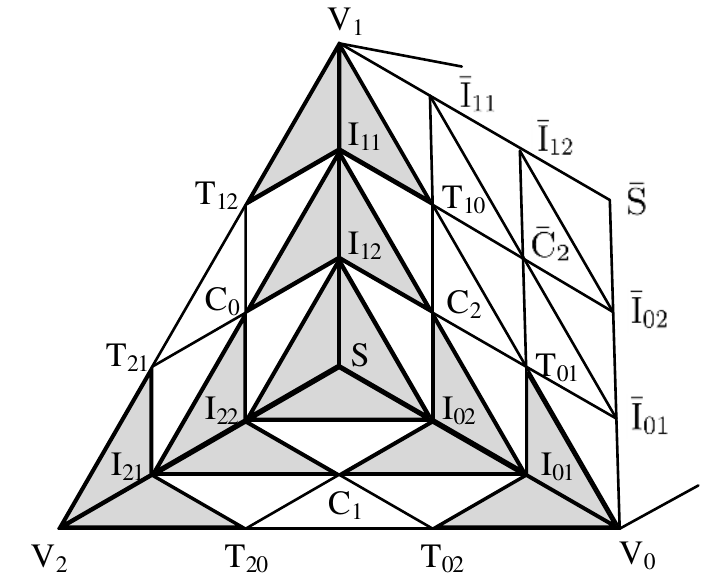}
  \caption{Top view of one triangle of the triangulation, showing its three microtriangles and its 19 B\'ezier ordinates.}\label{fig:cloughtocher}
\end{figure}

Our third assumption is that the quality measurement is precise. Because the HVS is the ultimate receiver in most applications, subjective evaluation is a straightforward and reliable approach to evaluate the quality of digital videos. Traditional subjective experiment protocol models a subject's perceived quality as a random variable, assuming the quality labeling process to be stochastic. Because subjective experiment is expensive and time consuming, it is hardly used in the GRD function approximation process. In practice, objective VQA methods that produce precise quality predictions are often employed to generate ground truth samples in the GRD function. Therefore, a GRD function should pass through the quality scores of objective VQA evaluated on the encoded video representations.

Under these assumptions, we define the space of GRD functions as:
\begin{align}
\nonumber W_{GRD} :=& \{f|f(x_n, y_n)=z_n, \forall n \in N, f \in C^1: \mathbb{R}^2 \to \mathbb{R}^J \\
\nonumber &\mathrm{and \ } \forall x_a < x_b, f(x_a, y) < f(x_b, y)\},
\end{align}
where $N$, $x_n$, $y_n$, and $z_n$ represent the total number of training samples, bitrate, spatial resolution, and quality of the $n$-th training sample, respectively.

In the subsequent sections, we introduce the proposed GRD model. Section~\ref{sec:ct_review} and~\ref{sec:monotonicity} review the traditional CT method and the monotonicity condition of cubic polynomial B\'ezier function, which the proposed model relies on. The proposed $C^1$ continuity condition, optimization framework, and robust axial-monotonic CT algorithm are novel contributions that are detailed in Section~\ref{sec:ct_affine},~\ref{sec:smoothness}, and~\ref{sec:robust}, respectively.

\subsection{Review of Clough-Tocher Method}\label{sec:ct_review}
Since first introduced in 1960's~\cite{clough1965ct}, the CT method has been the most widely used multi-dimensional scattered data interpolant, thanks to its $C^1$ continuity and low computational complexity~\cite{alfeld1984trivariate,amidror2002scattered}. Consider the scattered points $(x_n, y_n)$ located in the $x, y$ plane and their values $z_n$ over the plane, the triangulation of the scattered points in the $x, y$ plane induces a piecewise triangular surface over the plane, whose nodes are the points $(x_n, y_n, z_n)$. In CT method, a piecewise cubic function is employed as the interpolant for each triangle. Specifically, each triangle is further divided into three equivalent subtriangles, where a cubic function in the form of B\'ezier surface is estimated. Hereafter, we refer to the overall triangle as the macrotriangle and its subtriangles as microtriangles. The split triangular net is known as the control net, which is demonstrated in Fig.~\ref{fig:cloughtocher}. For clarity and brevity, we also denote the macrotriangle edge that is opposite to the vertex $V_i, i=0,1,2$ by $E_i$, and the internal microtriangle edge that connects $V_i$ and $S$ by $\hat{E}_i$. Mathematically, a cubic B\'ezier surface in the microtriangle $\Delta_{V_0V_1S}$ can be formulated as
\begin{align}\label{eq:bezier}
\nonumber z(\alpha,\beta,\gamma) =& c_{V_0}{\alpha}^3+3c_{T_{01}}{\alpha}^2{\beta}+3c_{I_{01}}{\alpha}^2{\gamma}+c_{V_1}{\beta}^3 +\\
\nonumber                         & 3c_{T_{10}}{\alpha}{\beta}^2+3c_{I_{11}}{\beta}^2{\gamma}+c_{S}{\gamma}^3+3c_{I_{02}}{\alpha}{\gamma}^2 +\\
                                  & 3c_{I_{12}}{\beta}{\gamma}^2 + 6c_{C_2}{\alpha}{\beta}{\gamma},
\end{align}
where $c_V$ represents the control net value at $V$ and $(\alpha,\beta,\gamma)$ is the barycentric coordinates with regard to the three vertices of the microtriangle. The barycentric coordinates of a point $P$ with regard to $\Delta_{V_0V_1S}$ can be defined as
\begin{align}
\nonumber \alpha_P = \frac{A_{PV_1S}}{A_{V_0V_1S}},
\beta_P = \frac{A_{PSV_0}}{A_{V_1SV_0}},
\gamma_P = \frac{A_{PV_0V_1}}{A_{SV_0V_1}},
\end{align}
where $A_{UVW}$ means the directional area of the triangle formed by points $U,V,W$ and is positive when $U,V,W$ is counter-clockwise. The conversion from Cartesian coordinate to barycentric coordinate is lengthy and thus omitted here. Interested readers may refer to~\cite{alfeld1984trivariate} for more details.

We note that 10 parameters are required to define a B\'ezier surface on each microtriangle. In the case of microtriangle $\Delta_{V_0V_1S}$, the parameters $c_{V_0}$, $c_{V_1}$, $c_{C_2}$ and $c_{S}$ are associated with point $V_0$, $V_1$, $C_2$ and $S$, respectively. The rest of the parameters are associated with the 6 trisection points on the three edges of $\Delta_{V_0V_1S}$. At the first glance, we need to determine 30 parameters for the interpolant in one triangle with only 3 equality constraints given by $c_{V_0}=z_{V_0}$, $c_{V_1}=z_{V_1}$, and $c_{V_2}=z_{V_2}$. Fortunately, the degree of freedom can be dramatically reduced with certain smoothness constraints. Under the $C^0$ assumption within macrotriangles, each two B\'ezier surfaces share the same curves at the their common boundaries $V_0S$, $V_1S$, and $V_2S$, leaving 19 free parameters in the macrotriangle $\Delta_{V_0V_1V_2}$. The inner-triangle $C^1$ continuity removes 7 additional degree of freedoms by enforcing the shaded neighboring microtriangles in Fig.~\ref{fig:cloughtocher} to be coplanar~\cite{farin1980bezier}. To ensure inter-triangle $C^1$ continuity, a standard approach is to assume the cross-boundary derivatives of the neighboring macrotriangles to be collinear, which further reduces the degree of freedom to 9. Taking into account the three known values at $V_0$, $V_1$, and $V_2$, we eventually have 6 unknown parameters in each macrotriangle. Although the gradients at vertices is not always available in practice, in most cases they can be estimated by considering the known values not only in the vertices of the triangle in question, but also in its neighbors. The most commonly used method is to estimate the gradients by minimizing the second-order derivatives along all B\'ezier curves~\cite{nielson1983method}. Readers who are interested in the details of the CT method may refer to~\cite{nielson1983method,farin1985modified,alfeld1984trivariate,amidror2002scattered}.

The original CT method suffers from at least three limitations in approximating GRD functions. First, it picks the normal of the edges as the direction of cross-boundary derivative $d^e_{E_i}$. However, this choice gives an interpolant that is not invariant under affine transforms. This has some undesirable consequences: for a very narrow triangle, the spline can develop huge oscillations~\cite{farin1985modified}. Second, the interpolant composite of piece-wise B\'ezier polynomials is not axial-monotonic, even when the given points are axial monotonic. Third, the CT algorithm achieves the external smoothness by estimating the gradients at three vertices $V_i, i=0,1,2$, and by assuming the normal derivative at the triangle boundary $E_i$ to be linear. The linear assumption is somewhat arbitrary and may violate monotonicity we want to achieve. We will address the three limitations in the subsequent sections.

\subsection{Affine-Invariant C1 Continuity}\label{sec:ct_affine}
In this section, we propose an affine invariant CT interpolant. Instead of the normal derivative at the triangle boundary $E_i$, we consider $d^e_{E_i}$ to be parallel to $\vec{C_i\bar{C}_i}$, \textit{i. e.}
\begin{subequations}\label{eq:innerC1}
  \begin{align}
  c_{P_i}=&(x_{P_i}-x_{V_i})d^x_{V_i}+(y_{P_i}-y_{V_i})d^y_{V_i}+z_{V_i} \label{eq:CTC1a} \\
  c_{C_i}=&\theta_{kj}c_{T_{jk}}+\theta_{jk}c_{T_{kj}}+\eta_id^e_{E_i} \label{eq:innerC1a}\\
  \nonumber c_{I_{i2}}=&\frac{1}{3}[(x_{I_{i1}}-x_{V_i})+(x_{T_{ki}}-x^*_j)+(x_{T_{ji}}-x^*_k)]d^x_{V_i} +\\
  \nonumber &\frac{1}{3}[(y_{I_{i1}}-y_{V_i})+(y_{T_{ki}}-y^*_j)+(y_{T_{ji}}-y^*_k)]d^y_{V_i} +\\
  \nonumber &\frac{1}{3}(x_{T_{ij}}-x^*_k)d^x_{V_j}+\frac{1}{3}(y_{T_{ij}}-y^*_k)d^y_{V_j} + \frac{1}{3}\eta_kd^e_{E_k} +\\
  \nonumber &\frac{1}{3}(x_{T_{ik}}-x^*_j)d^x_{V_k}+\frac{1}{3}(y_{T_{ik}}-y^*_j)d^y_{V_k} + \frac{1}{3}\eta_jd^e_{E_j} +\\
  &\frac{1}{3}[z_{V_i}+(\theta_{ki}z_{V_i}+\theta_{ik}z_{V_k})+(\theta_{ij}z_{V_j}+\theta_{ji}z_{V_i})] \label{eq:innerC1b}\\
  \nonumber c_S=&\frac{1}{9}\sum_{i=0}^{2}[(x_{I_{i1}}-x_{V_i})+2(x_{T_{ki}}-x^*_j)+2(x_{T_{ji}}-x^*_k)]d^x_{V_i} +\\
  \nonumber&\frac{1}{9}\sum_{i=0}^{2}[(y_{I_{i1}}-y_{V_i})+2(y_{T_{ki}}-y^*_j)+2(y_{T_{ji}}-y^*_k)]d^y_{V_i} +\\
  &\frac{2}{9}\sum_{i=0}^{2}\eta_id^e_{E_i}+\frac{1}{9}\sum_{i=0}^2[(1+2\theta_{ji}+2\theta_{ki})z_{V_i}], \label{eq:innerC1c}
  \end{align}
\end{subequations}
where
$$P_i \in \{T_{ij}, T_{ik}, I_{i1}\},$$
\resizebox{\hsize}{!}{$x^*_i=\frac{(x_{\bar{C}_i}-x_{C_i})(x_{V_j}y_{V_k}-x_{V_k}y_{V_j})-(x_{V_k}-x_{V_j})(x_{C_i}y_{\bar{C}_i}-x_{\bar{C}_i}y_{C_i})}{(x_{\bar{C}_i}-x_{C_i})(y_{V_k}-y_{V_j})-(y_{\bar{C}_i}-y_{C_i})(x_{V_k}-x_{V_j})},$}
\resizebox{\hsize}{!}{$y^*_i=\frac{(y_{\bar{C}_i}-y_{C_i})(x_{V_j}y_{V_k}-x_{V_k}y_{V_j})-(y_{V_k}-y_{V_j})(x_{C_i}y_{\bar{C}_i}-x_{\bar{C}_i}y_{C_i})}{(x_{\bar{C}_i}-x_{C_i})(y_{V_k}-y_{V_j})-(y_{\bar{C}_i}-y_{C_i})(x_{V_k}-x_{V_j})},$}
$$\eta_i=\sqrt{(x_{C_i}-x^*_i)^2+(y_{C_i}-y^*_i)^2},$$
$$\theta_{kj}=\frac{x_{T_{kj}}-x^*_i}{x_{T_{kj}}-x_{T_{jk}}},$$
$$\theta_{jk}=\frac{x_{T_{jk}}-x^*_i}{x_{T_{jk}}-x_{T_{kj}}},$$
$d^x_{V_i}$ and $d^y_{V_i}$ are partial derivatives of the B\'ezier surface at $V_i$ and $\{i, j, k\}$ is a cyclic permutation of $\{0, 1, 2\}$. Since this quantity transforms similarly as the gradient under affine transforms, the resulting interpolant is affine-invariant~\cite{farin1985modified}.

We also lift the unwanted linear constraints on the cross-boundary derivatives, elevating the number of parameters in a macrotriangle back to 9. In summary, the equality constraints in~\eqref{eq:innerC1} can be factorized into the matrix form for simplicity
\begin{equation}\label{eq:innerC1Mat}
  {\bf c}={\bf Rd} + {\bf f},
\end{equation}
where ${\bf c} \in \mathbb{R}^{16\times 1}$, ${\bf R} \in \mathbb{R}^{16\times 9}$, ${\bf d} \in \mathbb{R}^{9\times 1}$, ${\bf f} \in \mathbb{R}^{16\times 1}$, ${\bf c}$ and ${\bf d}$ represent the values of control net and unknown derivatives, respectively. Therefore, finding the interpolant of the macrotriangle corresponds to determining the 9 unknown parameters in ${\bf d}$.

Besides the inner macrotriangle constraints, we also want to keep $d^e_{E_i}$ consistent across the triangle boundary to ensure external $C^1$ smoothness. As a result, the following equality constraints need to be added for each edge with adjacent triangles
\begin{align}\label{eq:cbc1}
  d^e_{E_i}+d^e_{\bar{E}_i}=0.
\end{align}

Combining~\eqref{eq:innerC1Mat} and~\eqref{eq:cbc1}, we conclude that the resulting function is $C^1$ continuous and affine-invariant.

\subsection{Axial Monotonicity}\label{sec:monotonicity}
This section aims to derive the sufficient constraints on ${\bf d}$ for the B\'ezier surface in the macrotriangle $\Delta_{V_0V_1V_2}$ to be axial-monotonic. In general, the interpolant composite of piece-wise B\'ezier polynomials is not monotonic even though the sampled points are monotonic. Several works have been done to derive sufficient conditions for a univariate or bivariate B\'ezier function~\cite{fritsch1980monotone,han1997fitting}. We adopt the sufficient condition proposed in~\cite{han1997fitting}, where it was proved that the cubic B\'ezier surface in a microtriangle is axial-monotonic when all the 6 triangular patches of its control net (\textit{e.g.} $\Delta_{I_{02}I_{12}S}$, $\Delta_{C_{2}I_{11}I_{12}}$, $\Delta_{C_{2}I_{01}I_{02}}$, $\Delta_{V_{1}T_{10}I_{11}}$, $\Delta_{T_{10}T_{01}C_{2}}$, and $\Delta_{T_{01}V_{0}I_{01}}$ in $\Delta_{V_0V_1S}$) are axial-monotonic. By combining the sufficient conditions in all three microtriangles and the inner triangle continuity, we obtain 
\begin{subequations}\label{eq:mono}
  \begin{align}
  &\resizebox{0.8\hsize}{!}{$(y_{V_i}-y_{V_k})c_{T_{ij}}+(y_{V_j}-y_{V_i})c_{T_{ik}} \leq (y_{V_j}-y_{V_k})z_{V_i}$} \label{eq:monoA} \\
  &\resizebox{0.8\hsize}{!}{$(y_{V_k}-y_{V_j})c_{I_{i1}}+(y_{V_i}-y_{V_k})c_{C_k}+(y_{V_j}-y_{V_i})c_{C_j} \leq 0$} \label{eq:monoB} \\
  &\resizebox{0.8\hsize}{!}{$(y_{V_2}-y_{V_1})c_{I_{02}}+(y_{V_0}-y_{V_2})c_{I_{12}}+(y_{V_1}-y_{V_0})c_{I_{22}} \leq 0$} \label{eq:monoC} \\
  &\resizebox{0.8\hsize}{!}{$(y_{S}-y_{V_j})c_{T_{ij}}+(y_{V_i}-y_{S})c_{T_{ji}}+(y_{V_j}-y_{V_i})c_{C_k} \leq 0$}. \label{eq:monoD}
  \end{align}
\end{subequations}
We can summarize the monotonicity constraint in matrix form
\begin{equation}\label{eq:monoMat}
  {\bf Gc} \leq {\bf h},
\end{equation}
where ${\bf G} \in \mathbb{R}^{10\times 16}$ and ${\bf f} \in \mathbb{R}^{10\times 1}$. Further substitute~\eqref{eq:innerC1Mat} into~\eqref{eq:monoMat}, we obtain the monotonicity constraint in terms of ${\bf d}$
\begin{equation}\label{eq:monoMatD}
  {\bf GRd} \leq {\bf h-Gf}.
\end{equation}
More details on how we construct $\bf G$ and $\bf h$ are given in the Appendix.

\subsection{Optimization-based Solutions}\label{sec:smoothness}
To determine the unknown derivatives, we propose to minimize the total curvature of the interpolated surface under the smoothness assumption. Directly computing the total curvature is computationally intractable. Alternatively, we minimize the curvature of B\'ezier curves at the edges of each microtriangle as its approximation. Specifically, in $\Delta_{V_0V_1V_2}$, the objective function is written as

\begin{align}\label{eq:loss}
\resizebox{0.9\hsize}{!}{$L_{V_0V_1V_2}=\frac{1}{2}\sum_{i=0}^{2}\int_{E_i}\left[\frac{\partial^2 z}{\partial E_i^2}\right]^2ds_{E_i} + \sum_{i=0}^{2}\int_{\hat{E}_i}\left[\frac{\partial^2 z}{\partial \hat{E}_i^2}\right]^2ds_{\hat{E}_i}$},
\end{align}
where the weight $\frac{1}{2}$ is introduced to cancel the double counting of the external edges.

Consider an external boundary $E_i$, whose B\'ezier control net coefficients are $z_{V_j}$, $c_{T_{jk}}$, $c_{T_{kj}}$, and $z_{V_k}$. The integral of the second order derivative of the B\'ezier curve on $E_i$ can be represented in terms of the four coefficients as

\begin{align}\label{eq:loss1}
\nonumber&\int_{E_i}\left[\frac{\partial^2 z}{\partial E_i^2}\right]^2ds_{E_i}=\frac{1}{\|E_i\|^3}\int_0^1\left[z_{E_i}^{''}(t)\right]^2dt \\
\nonumber=&\frac{18}{\|E_i\|^3}(2c_{T_jk}^2+2c_{T_kj}^2-2c_{T_jk}c_{T_kj})+ \\
\nonumber&\frac{-36}{\|E_i\|^3}(z_{V_j}c_{T_jk}+z_{V_k}c_{T_kj}) + \frac{12}{\|E_i\|^3}(z_{V_j}^2+z_{V_k}^2+z_{V_j}z_{V_k}) \\
\nonumber=&\left[ {\begin{array}{*{20}{c}}
{{c_{{T_{jk}}}}}&{{c_{{T_{kj}}}}}
\end{array}} \right]\left[ {\begin{array}{*{20}{c}}
{\frac{{36}}{{{{\left\| {{E_i}} \right\|}^3}}}}&{\frac{{ - 18}}{{{{\left\| {{E_i}} \right\|}^3}}}}\\
{\frac{{ - 18}}{{{{\left\| {{E_i}} \right\|}^3}}}}&{\frac{{36}}{{{{\left\| {{E_i}} \right\|}^3}}}}
\end{array}} \right]\left[ {\begin{array}{*{20}{c}}
{{c_{{T_{jk}}}}}\\
{{c_{{T_{kj}}}}}
\end{array}} \right] + \\
\nonumber&\left[ {\begin{array}{*{20}{c}}
{\frac{{ - 36{z_{{V_j}}}}}{{{{\left\| {{E_i}} \right\|}^3}}}}&{\frac{{ - 36{z_{{V_k}}}}}{{{{\left\| {{E_i}} \right\|}^3}}}}
\end{array}} \right]\left[ {\begin{array}{*{20}{c}}
{{c_{{T_{jk}}}}}\\
{{c_{{T_{kj}}}}}
\end{array}} \right] +\\
&\frac{{12}}{{{{\left\| {{E_i}} \right\|}^3}}}(z_{{V_j}}^2 + z_{{V_k}}^2 + {z_{{V_j}}}{z_{{V_k}}}),
\end{align}
where
\begin{align}
\nonumber \|E_i\|=\sqrt{(x_{V_j}-x_{V_k})^2+(y_{V_j}-y_{V_k})^2}
\end{align}
is the length of $E_i$.

Similarly, we get the other part of the objective function from an internal boundary $\hat{E}_i$, whose coefficients are $z_{V_i}, c_{I_{i1}}, c_{I_{i2}},$ and $c_{S}$.

\begin{align}\label{eq:loss2}
\nonumber&\int_{\hat{E}_i}\left[\frac{\partial^2 z}{\partial \hat{E}_i^2}\right]^2ds_{\hat{E}_i}=\frac{1}{\|\hat{E}_i\|^3}\int_0^1\left[z_{\hat{E}_i}^{''}(t)\right]^2dt \\
\nonumber=&\frac{6}{\|\hat{E}_i\|^3}(6c_{I_{i1}}^2+6c_{I_{i2}}^2+2c_S^2-6c_{I_{i1}}c_{I_{i2}}-6c_{I_{i2}}c_S)+ \\
\nonumber&\frac{12z_{V_i}}{\|\hat{E}_i\|^3}(-3c_{I_{i1}}+c_S)+\frac{12}{\|\hat{E}_i\|^3}z_{V_i}^2 \\
\nonumber=&\left[ {\begin{array}{*{20}{c}}
{{c_{{I_{i1}}}}}&{{c_{{I_{i2}}}}}&{{c_S}}
\end{array}} \right]\left[ {\begin{array}{*{20}{c}}
{\frac{{36}}{{{{\left\| {{{\hat E}_i}} \right\|}^3}}}}&{\frac{{ - 18}}{{{{\left\| {{{\hat E}_i}} \right\|}^3}}}}&{0}\\
{\frac{{ - 18}}{{{{\left\| {{{\hat E}_i}} \right\|}^3}}}}&{\frac{{36}}{{{{\left\| {{{\hat E}_i}} \right\|}^3}}}}&{\frac{{ - 18}}{{{{\left\| {{{\hat E}_i}} \right\|}^3}}}}\\
{0}&{\frac{{ - 18}}{{{{\left\| {{{\hat E}_i}} \right\|}^3}}}}&{\frac{{12}}{{{{\left\| {{{\hat E}_i}} \right\|}^3}}}}
\end{array}} \right]\left[ {\begin{array}{*{20}{c}}
{{c_{{I_{i1}}}}}\\
{{c_{{I_{i2}}}}}\\
{{c_S}}
\end{array}} \right] \\
& + \left[ {\begin{array}{*{20}{c}}
{\frac{{ - 36{z_{{V_i}}}}}{{{{\left\| {{{\hat E}_i}} \right\|}^3}}}}&0&{\frac{{12{z_{{V_i}}}}}{{{{\left\| {{{\hat E}_i}} \right\|}^3}}}}
\end{array}} \right]\left[ {\begin{array}{*{20}{c}}
{{c_{{I_{i1}}}}}\\
{{c_{{I_{i2}}}}}\\
{{c_S}}
\end{array}} \right] + \frac{{12z_{V_i^{}}^2}}{{{{\left\| {{{\hat E}_i}} \right\|}^3}}},
\end{align}
where
\begin{align}
\nonumber \|\hat{E}_i\|=\sqrt{(x_{S}-x_{V_i})^2+(y_{S}-y_{V_i})^2}
\end{align}
is the length of $\hat{E}_i$.

Substitute~\eqref{eq:loss1} and~\eqref{eq:loss2} into~\eqref{eq:loss}, we obtain the loss function for $\Delta_{V_0V_1V_2}$ in matrix form

\begin{align}\label{eq:loss3}
  L_{V_0V_1V_2}={\bf c}^T{\bf U}_{V_0V_1V_2}{\bf c}+{\bf w}_{V_0V_1V_2}^T{\bf c} + const,
\end{align}
where ${\bf U}_{V_0V_1V_2} \in \mathbb{R}^{16\times 16}$ and ${\bf w}_{V_0V_1V_2}\in \mathbb{R}^{16\times 1}$.

Substituting ${\bf c} = {\bf Rd}+{\bf f}$ into~\eqref{eq:loss3}, we get
\begin{align}\label{eq:obj}
\nonumber L_{V_0V_1V_2}=&({\bf Rd}+{\bf f})^T{\bf U}_{V_0V_1V_2}({\bf Rd}+{\bf f}) + \\
\nonumber               &{\bf w}_{V_0V_1V_2}^T({\bf Rd}+{\bf f}) + const \\
\nonumber =&{\bf d}^T({\bf R}^T{\bf U}_{V_0V_1V_2}{\bf R}){\bf d} + \\
 &({\bf f}^T{\bf U}_{V_0V_1V_2}+{\bf w}_{V_0V_1V_2}^T){\bf Rd} + const.
\end{align}

In summary, finding the axial-monotonic interpolant corresponds to solving the following optimization problem
\begin{equation}
\resizebox{\hsize}{!}{$
\begin{aligned}\label{eq:opt}
& \underset{{\bf d}}{\text{minimize}}
& & {\bf d}^T({\bf R}^T{\bf U}_{V_0V_1V_2}{\bf R}){\bf d} + ({\bf f}^T{\bf U}_{V_0V_1V_2}+{\bf w}_{V_0V_1V_2}^T){\bf Rd} \\
& \text{subject to}
& & {\bf G}{\bf R}{\bf d} \leq {\bf h} - {\bf G}{\bf f}, \\
&&& d^e_{E_i}+d^e_{\bar{E}_i}=0.
\end{aligned}
$}
\end{equation}
Note that the constraints are linear with respect to ${\bf d}$ and ${\bf R}^T{\bf U}_{V_0V_1V_2}{\bf R}$ is positive-semidefinite. Thus, finding ${\bf d}$ turns into a standard problem of quadratic programming, which can be efficiently solved by the existing convex programming packages~\cite{stellato2017osqp}.

\subsection{Robust Axial-Monotonic Clough-Tocher Method}\label{sec:robust}
Here we propose our Robust Axial-Monotonic Clough-Tocher (RAMCT) method. The inequality constraints in~\eqref{eq:monoMat} are sufficient conditions for $x$-axial monotonicity. However, the sufficient conditions cannot be satisfied in some extreme cases, making the primary solution infeasible. To relax these constraints, we introduce hinge loss to some of these inequalities, motivated by the success of Support Vector Machine~\cite{cortes1995svm}. Specifically, the modified inequality constraints are formulated as
\begin{subequations}\label{eq:relax}
\begin{align}
& \resizebox{0.85\hsize}{!}{$(y_{V_i}-y_{V_k})c_{T_{ij}}+(y_{V_j}-y_{V_i})c_{T_{ik}} \leq (y_{V_j}-y_{V_k})z_{V_i}$} \label{eq:relaxA} \\
& \resizebox{0.85\hsize}{!}{$(y_{V_k}-y_{V_j})c_{I_{i1}}+(y_{V_i}-y_{V_k})c_{C_k}+(y_{V_j}-y_{V_i})c_{C_j}+\xi_{i1} \leq 0 $} \label{eq:relaxB} \\
& \resizebox{0.85\hsize}{!}{$(y_{V_2}-y_{V_1})c_{I_{02}}+(y_{V_0}-y_{V_2})c_{I_{12}}+(y_{V_1}-y_{V_0})c_{I_{22}} \leq 0 $} \label{eq:relaxC} \\
& \resizebox{0.85\hsize}{!}{$(y_{S}-y_{V_j})c_{T_{ij}}+(y_{V_i}-y_{S})c_{T_{ji}}+(y_{V_j}-y_{V_i})c_{C_k}+\xi_{C_k} \leq 0, $} \label{eq:relaxD}
\end{align}
\end{subequations}
where $\boldsymbol \xi \in \mathbb{R}^{6\times 1}$ and $\boldsymbol \xi \leq {\bf 0}$. Note that~\eqref{eq:relaxA},\eqref{eq:relaxC} are identical to~\eqref{eq:monoA},\eqref{eq:monoC} because they are necessary conditions of axial monotonicity (See Appendix for proof). Rewriting these constraints in the matrix form, we obtain

\begin{align}
\nonumber \left[ {\begin{array}{*{20}{c}}
{\bf{G}}&{\bf{J}}_1\\
{\bf{O}}&{\bf{J}_2}
\end{array}} \right]\left[ \begin{array}{l}
{\bf c}\\
\boldsymbol \xi 
\end{array} \right] \le \left[ \begin{array}{l}
{\bf{h}}\\
{\bf{0}}
\end{array} \right],
\end{align}
where ${\bf G}$ and ${\bf h}$ are the same as in~\eqref{eq:monoMat},\eqref{eq:monoMatD}. ${\bf J}_2$ is a $6\times 6$ identity matrix, while ${\bf J}_1 \in \mathbb{R}^{10\times 6}$ is obtained by padding ${\bf J}_2$ with 3 rows of zeros to its top and inserting a row of zeros between the 3rd and 4th rows of ${\bf J}_2$.

By substituting~\eqref{eq:innerC1Mat} into the inequality above, we finally obtain the inequality constraints in terms of the unknowns ${\bf d}$ and the auxiliary variables ${\boldsymbol \xi}$ as
\begin{align}\label{eq:robustMono}
&\left[ {\begin{array}{*{20}{c}}
{{\bf{GR}}}&{{{\bf{J}}_1}}\\
{\bf{O}}&{{{\bf{J}}_2}}
\end{array}} \right]\left[ \begin{array}{l}
{\bf{d}}\\
{\boldsymbol {\xi }}
\end{array} \right] \le \left[ \begin{array}{l}
{\bf{h}} - {\bf{Gf}}\\
{\bf{0}}
\end{array} \right].
\end{align}
The objective function is then modified accordingly,
\begin{align}\label{eq:robust}
L_{V_0V_1V_2}={\bf c}^T{\bf U}_{V_0V_1V_2}{\bf c}+{\bf w}_{V_0V_1V_2}^T{\bf c} - \boldsymbol{\lambda}^T\boldsymbol\xi + const,
\end{align}
where $\boldsymbol{\lambda} = [\lambda, \lambda, \cdots, \lambda]^T$ is the weighting parameter. Substituting ${\bf c} = {\bf Rd}+{\bf f}$ into~\eqref{eq:robust}, we get

\begin{align}\label{eq:robustObj}
\nonumber &L_{V_0V_1V_2}={\bf d}^T({\bf R}^T{\bf U}_{V_0V_1V_2}{\bf R}){\bf d} + ({\bf f}^T{\bf U}_{V_0V_1V_2}+\\
\nonumber& {\bf w}_{V_0V_1V_2}^T){\bf Rd} - \boldsymbol{\lambda}^T\boldsymbol\xi +const \\
\nonumber =&\left[ {\begin{array}{*{20}{c}}
{{{\bf{d}}^T}}&{{{\boldsymbol{\xi }}^T}}
\end{array}} \right]\left[ {\begin{array}{*{20}{c}}
{{{\bf{R}}^{\bf{T}}}{{\bf{U}}_{{V_0}{V_1}{V_2}}}{\bf{R}}}&\bf{O}\\
\bf{O}&\bf{O}
\end{array}} \right]\left[ {\begin{array}{*{20}{c}}
{\bf{d}}\\
{\boldsymbol{\xi }}
\end{array}} \right] + \\
& \left[ {\begin{array}{*{20}{c}}
{({{\bf{f}}^T}{{\bf{U}}_{{V_0}{V_1}{V_2}}} + {\bf{w}}_{{V_0}{V_1}{V_2}}^T){\bf{R}}}&{ - {{\boldsymbol{\lambda }}^T}}
\end{array}} \right]\left[ {\begin{array}{*{20}{c}}
{\bf{d}}\\
{\boldsymbol{\xi }}
\end{array}} \right] + const.
\end{align}
Replacing~\eqref{eq:obj},\eqref{eq:monoMat} with~\eqref{eq:robustObj},\eqref{eq:robustMono} in~\eqref{eq:opt}, we find that the original interpolation problem remains to be a quadratic programming problem.

\section{Information-theoretic sampling}\label{sec:sampling}
In this section, we first explore the informativeness of samples in the GRD space via a probabilistic model. We then present an information-theoretic sampling strategy that optimally selects the samples, offering enormous savings in time and computational resources.

\begin{figure}[!t]
  \removelatexerror
  \begin{algorithm}[H]
   \caption{Uncertainty Sampling}
   Initialize $S = \emptyset$; $\boldsymbol{\bar{\Sigma}}^{(1)}=\boldsymbol{\Sigma}$ \;
   \For{$k := 1$ to $K$}
   {
      $i^{(k)} = \underset{i}{\text{minimize}} \quad \mathrm{tr} (\boldsymbol{\bar{\Sigma}}_{ii}^{(k)} - \frac{{\boldsymbol{\bar{\sigma}}_i^{(k)}}^T\boldsymbol{\bar{\sigma}}_i^{(k)}}{\bar{\sigma}_{ii}^{(k)}})$ \;
      $x^{(k)}$ = VQA(Encode(${\bf r}_i^{(k)}$))\;
      Set $S = S \cup x^{(k)}$ \;
      $\boldsymbol{\bar{\Sigma}}^{(k+1)} = \boldsymbol{\bar{\Sigma}}_{ii}^{(k)} - \frac{{\boldsymbol{\bar{\sigma}}_i^{(k)}}^T\boldsymbol{\bar{\sigma}}_i^{(k)}}{\bar{\sigma}_{ii}^{(k)}}$ \;
      \If{$\mathrm{tr}(\boldsymbol{\bar{\Sigma}}_{ii}^{(k)} - \frac{{\boldsymbol{\bar{\sigma}}_i^{(k)}}^T\boldsymbol{\bar{\sigma}}_i^{(k)}}{\bar{\sigma}_{ii}}) \leq T$}{
        Break \;
      }
   }
  \end{algorithm}
\end{figure}

Let ${\bf x}=(x_1, ..., x_N)$ be a vector of discrete samples on a GRD function uniformly distributed in the bitrate-resolution space, where $N$ is the total number of sample points on the grid. Given that the GRD function is smooth, when the sampling grid is dense, these discrete samples provide a good description of the continuous GRD function. In particular, when the GRD function is band-limited, it can be fully recovered from these samples when the sampling density is larger than the Nyquist rate. Assuming \textbf{x} is created from GRD functions of real-world video content, we model \textbf{x} as an $N$-dimensional random variable, for which the probability density function $p_{\bf x}(\bf x) \sim \mathcal{N}(\boldsymbol{\mu},\,\boldsymbol{\Sigma})$ follows a multivariate Normal distribution. The total uncertainty of \textbf{x} is characterized by its joint entropy given by
\begin{align}\label{eq:ent}
&H_{\bf x}({\bf x}) = \frac{1}{2}\log|\boldsymbol{\Sigma}| + const,
\end{align}
where $|\cdot|$ is the determinant operator. If the full vector \textbf{x} is further divided into two parts such that ${\bf x} = \left[ {\begin{array}{*{20}{c}}
{\bf x}_1\\
{\bf x}_2
\end{array}} \right]$ and $\boldsymbol{\Sigma} = \begin{bmatrix}\boldsymbol{\Sigma}_{11} & \boldsymbol{\Sigma}_{12} \\\boldsymbol{\Sigma}_{21} & \boldsymbol{\Sigma}_{22} \end{bmatrix}$, and the ${\bf x}_2$ portion has been resolved by ${\bf x}_2={\bf a}$, then the remaining uncertainty is given by the conditional entropy
\begin{align}\label{eq:entropy}
&H_{{\bf x}_1|{\bf x}_2}({\bf x}_1|{\bf x}_2 = {\bf a}) = \frac{1}{2}\log|\boldsymbol{\bar{\Sigma}}|) + const,
\end{align}
where
\begin{align}\label{eq:conditionalVar}
&\boldsymbol{\bar{\Sigma}} = \boldsymbol{\Sigma}_{11} - \boldsymbol{\Sigma}_{12}\boldsymbol{\Sigma}_{22}^{-1}\boldsymbol{\Sigma}_{21}.
\end{align}
As a special case, we aim to find one sample that most efficiently reduces the uncertainty of GRD estimation. This is found by minimizing the log determinant of the conditional covariance matrix~\cite{bishop2006pattern}
\begin{equation}
\begin{aligned}\label{eq:sampling}
& \underset{i}{\text{minimize}}
& \log |\boldsymbol{\bar{\Sigma}}| = \underset{i}{\text{minimize}}
& & \log |\boldsymbol{\bar{\Sigma}}_{ii} - \frac{\boldsymbol{\bar{\sigma}}_i^T\boldsymbol{\bar{\sigma}}_i}{\bar{\sigma}_{ii}}|,
\end{aligned}
\end{equation}
where $\boldsymbol{\bar{\Sigma}} = \begin{bmatrix}\boldsymbol{\bar{\Sigma}}_{ii} & \boldsymbol{\bar{\sigma}}_i \\\boldsymbol{\bar{\sigma}}_i^T & \bar{\sigma}_{ii} \end{bmatrix}$ and $i$ is the row index of $\boldsymbol{\bar{\Sigma}}$.

\begin{figure}
  \centering
  \includegraphics[width=0.45\textwidth]{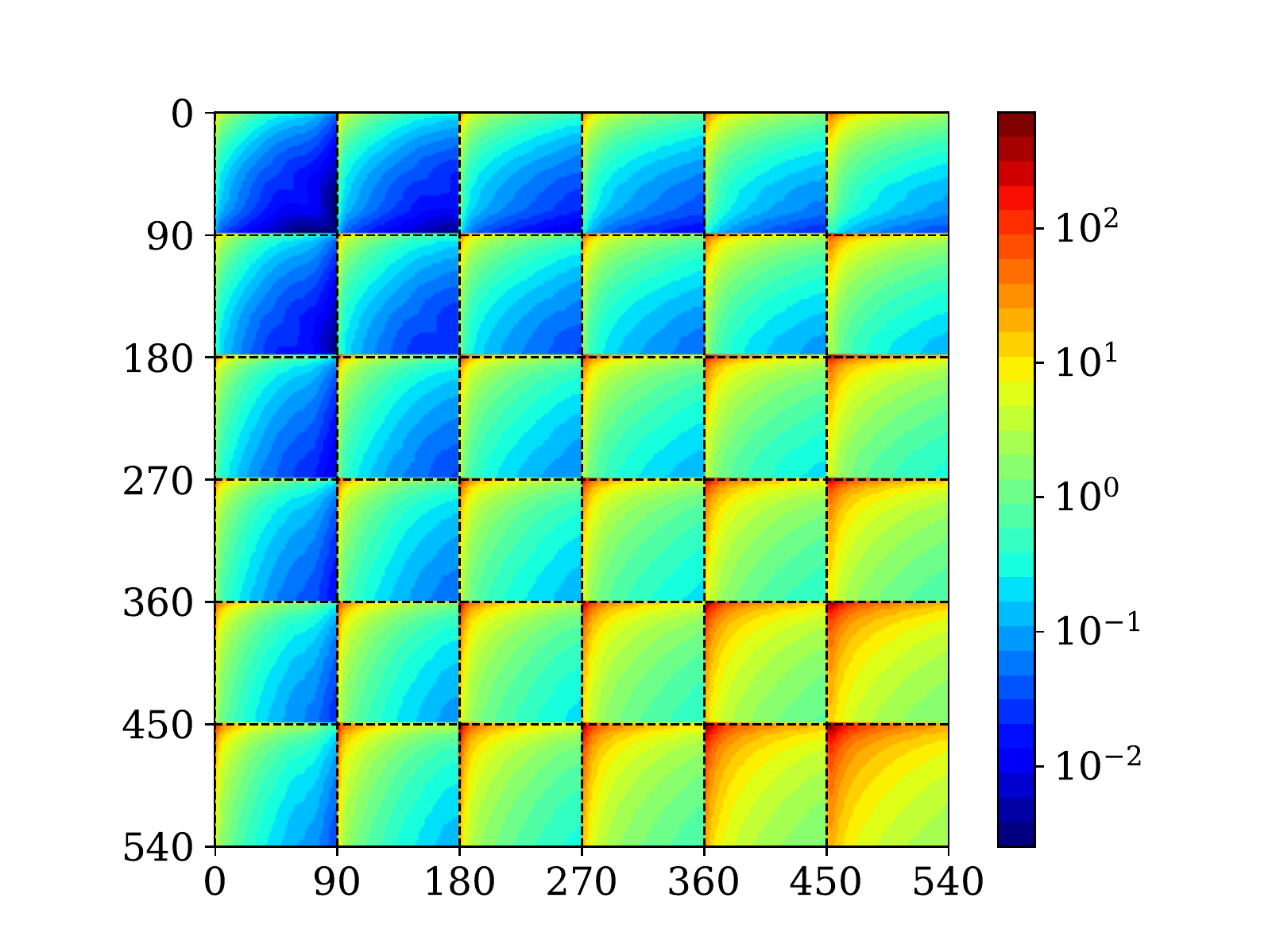}
  \caption{Empirical covariance matrix of the GRD functions. Bitrate and spatial resolution are presented in ascending order, and spatial resolution elevates every 90 samples.}\label{fig:cov}
\end{figure}

Minimizing~\eqref{eq:sampling} directly is computationally expensive, especially when the dimensionality is high. Alternatively, we minimize the upper bound of the conditional entropy
\begin{equation}
\begin{aligned}\label{eq:samplingTrace}
& \underset{i}{\text{minimize}}
& & \mathrm{tr}(\boldsymbol{\bar{\Sigma}}_{ii} - \frac{\boldsymbol{\bar{\sigma}}_i^T\boldsymbol{\bar{\sigma}}_i}{\bar{\sigma}_{ii}}),
\end{aligned}
\end{equation}
where $\log|\boldsymbol{\bar{\Sigma}}_{ii} - \frac{\boldsymbol{\bar{\sigma}}_i^T\boldsymbol{\bar{\sigma}}_i}{\bar{\sigma}_{ii}}| \leq \mathrm{tr}(\boldsymbol{\bar{\Sigma}}_{ii} - \frac{\boldsymbol{\bar{\sigma}}_i^T\boldsymbol{\bar{\sigma}}_i}{\bar{\sigma}_{ii}} - {\bf I})$ and ${\bf I}$ denotes identity matrix. The sample with the minimum average loss in~\eqref{eq:samplingTrace} over all viewing devices is most informative. Once the optimal sample index is obtained, we encode the video at the $i$-th representation, evaluate its quality with objective VQA algorithms, and update the conditional covariance matrix in~\eqref{eq:conditionalVar}. The process is applied iteratively until the overall uncertainty in the system is reduced below a certain threshold $T$. We summarize the proposed uncertainty sampling method in Algorithm 1, where ${\bf r}_i$ represents the bitrate and spatial resolution at the $i$-th representation.

\textbf{Remark}: To get a sense of what type of samples will be chosen by the proposed algorithm, we analyze several influencing factors in the objective function~\eqref{eq:samplingTrace}:

\begin{itemize}
  \item By the basic properties of trace, the objective function in the uncertainty sampling can be factorized as
  \begin{align}\label{eq:refactor}
  \nonumber &\mathrm{tr}(\boldsymbol{\bar{\Sigma}}_{ii} - \frac{\boldsymbol{\bar{\sigma}}_i^T\boldsymbol{\bar{\sigma}}_i}{\bar{\sigma}_{ii}}) \\
  \nonumber=& \mathrm{tr}(\boldsymbol{\bar{\Sigma}}_{ii}) - \frac{\mathrm{tr}(\boldsymbol{\bar{\sigma}}_i^T\boldsymbol{\bar{\sigma}}_i)}{\bar{\sigma}_{ii}}\\
  \nonumber=& \mathrm{tr}(\boldsymbol{\bar{\Sigma}}) - (\bar{\sigma}_{ii} + \frac{1}{\bar{\sigma}_{ii}}\sum\limits_{j\neq i} \bar{\sigma}_{ij}^2).
  \end{align}
Thus, $\mathrm{tr}(\boldsymbol{\bar{\Sigma}}_{ii} - \frac{\boldsymbol{\bar{\sigma}}_i^T\boldsymbol{\bar{\sigma}}_i}{\bar{\sigma}_{ii}})$ is a decreasing function with respect to $\bar{\sigma}_{ii}$ when $\bar{\sigma}_{ii} > \sqrt{\sum\limits_{j\neq i}\bar{\sigma}_{ij}^2}$. This indicates that samples with large uncertainty are more likely to be selected than those with small uncertainty.
  \item According to~\eqref{eq:conditionalVar}, $\forall j \neq i$,
  \begin{align}
  \nonumber & \bar{\sigma}_{jj}^{(k+1)} = \bar{\sigma}_{jj}^{(k)} - \frac{\bar{\sigma}_{ij}^{(k)^2}}{\bar{\sigma}_{ii}^{(k)}},
  \end{align}
  suggesting the rate of reduction in the uncertainty of sample $j$ is proportional to its squared correlation with the selected sample $i$ in the $k$-th iteration. Fig.~\ref{fig:cov} shows an empirical covariance matrix $\boldsymbol{\bar{\Sigma}}$ estimated from our video dataset that will be detailed in the next section, from which we observe that the GRD functions typically exhibit high correlation in a local region. Combining the first observation above, we conclude that the next optimal choice of sample should be selected from the region where labeled samples are sparse.
  \item Note that knowing that ${\bf x}_2 = {\bf a}$ alters the variance, though the new variance does not depend on the specific value of ${\bf a}$. The independence has two important consequences. First, the proposed sampling scheme is general enough to accommodate GRD estimators from all classes. More importantly, the algorithm results in a unique sampling sequence for all GRD functions. In other words, we can generate a lookup table of optimal querying order, making the sampling process fully parallelizable.
\end{itemize}


\begin{table*}
  \centering
  \caption{MSE performance of the competing GRD function models with different number of labeled samples selected by random sampling (RS) and the proposed uncertainty sampling (US).}\label{tab:mse}
  \begin{tabular}{c|C{1cm}C{1cm}|C{1cm}C{1cm}|C{1cm}C{1cm}|C{1cm}C{1cm}|C{1cm}C{1cm}}
      \toprule
     \multirow{2}{*}{sample \#}   & \multicolumn{2}{c|}{Reciprocal~\cite{kreuzberger2016comparative}} & \multicolumn{2}{c|}{Logarithmic~\cite{chen2016subjective}} & \multicolumn{2}{c|}{PCHIP}      & \multicolumn{2}{c|}{CT} & \multicolumn{2}{c}{RAMCT} \\ \cline{2-11}
                                  & RS & US & RS & US & RS & US & RS & US & RS & US \\ \hline
     20                           & N.A. & N.A. & 23.07 & \textbf{13.33} & 68.76 & 26.49 & 88.54 & 56.04 & 135.27 & 16.10 \\
     30                           & 62.27 & 83.34 & 13.08 & 10.56 & 30.95 & \textbf{2.06} & 37.99 & 22.78 & 10.98 & 3.29 \\
     50                           & 38.11 & 73.88 & 9.43  & 6.77 & 8.64 & 0.07 & 11.75 & 12.16 & 4.70 & \textbf{0.06} \\
     75                           & 30.27 & 48.85 & 5.15  & 4.92 & 3.08  & \textbf{0} & 4.84 & 3.26 & 1.01 & \textbf{0} \\
     100                          & 27.44 & 38.46 & 4.60  & 4.18 & 1.77  & \textbf{0} & 2.75  & 1.26 & 0.13 & \textbf{0} \\
     540                          & 24.51 & 24.51 & 2.76  & 2.76 & \textbf{0} & \textbf{0} & \textbf{0} & \textbf{0} & \textbf{0} & \textbf{0} \\
     \bottomrule
   \end{tabular}
\end{table*}

\begin{table*}
  \centering
  \caption{$l_{\infty}$ performance of the competing GRD function models with different number of labeled samples selected by random sampling (RS) and the proposed uncertainty sampling (US).}\label{tab:linf}
  \begin{tabular}{c|C{1cm}C{1cm}|C{1cm}C{1cm}|C{1cm}C{1cm}|C{1cm}C{1cm}|C{1cm}C{1cm}}
      \toprule
     \multirow{2}{*}{sample \#}   & \multicolumn{2}{c|}{Reciprocal~\cite{kreuzberger2016comparative}} & \multicolumn{2}{c|}{Logarithmic~\cite{chen2016subjective}} & \multicolumn{2}{c|}{PCHIP}      & \multicolumn{2}{c|}{CT} & \multicolumn{2}{c}{RAMCT} \\ \cline{2-11}
                                  & RS & US & RS & US & RS & US & RS & US & RS & US \\ \hline
     20                           & N.A. & N.A. & 19.40 & \textbf{16.56} & 38.87 & 28.11 & 36.50 & 29.51 & 45.15 & 21.88 \\
     30                           & 48.32 & 45.36 & 17.85 & 12.28 & 33.04 & 11.07 & 29.84 & 18.70 & 27.07 & \textbf{6.13} \\
     50                           & 52.48 & 45.48 & 15.75 & 12.37 & 24.33 & \textbf{2.10} & 21.82 & 14.30 & 23.99 & 2.13 \\
     75                           & 54.49 & 49.08 & 14.59 & 13.53 & 18.22 & 0.47 & 17.89 & 7.76 & 16.51 & \textbf{0.11} \\
     100                          & 55.54 & 51.26 & 14.22 & 14.44 & 16.00 & 0.26 & 15.59 & 5.84 & 14.23 & \textbf{0} \\
     540                          & 58.04 & 58.04 & 18.33 & 18.14 & \textbf{0} & \textbf{0} & \textbf{0} & \textbf{0} & \textbf{0} & \textbf{0} \\
     \bottomrule
   \end{tabular}
\end{table*}

\section{Experiments}\label{sec:performance}
In this section, we first describe the experimental setups including our GRD function database, the implementation details of the proposed algorithm, and the evaluation criteria. We then compare the proposed algorithm with existing GRD estimation methods.

\subsection{Experimental setups}\label{subsec:dataset}
\textbf{GRD Function Database:} We construct a new video database which contains 250 pristine videos that span a great diversity of video content. An important consideration in selecting the videos is that they need to be representative of the videos we see in the daily life. Therefore, we resort to the Internet and elaborately select 200 keywords to search for creative common licensed videos. We initially obtain more than 700 4K videos. Many of these videos contain significant distortions, including heavy compression artifacts, noise, blur, and other distortions due to improper operations during video acquisition and sharing. To make sure that the videos are of pristine quality, we carefully inspect each of the videos multiple times by zooming in and remove those videos with visible distortions. We further reduce artifacts and other unwanted contaminations by downsampling the videos to a size of 1920 $\times$ 1080 pixels, from which we extract 10 seconds semantically coherent video clips. Eventually, we end up with 250 high quality videos.

Using the aforementioned sequences as the source, each video is distorted by the following processes sequentially:
\begin{itemize}
  \item Spatial downsample: We downsample source videos using bi-cubic filter to six spatial resolutions (1920 $\times$ 1080, 1280 $\times$ 720, 720 $\times$ 480, 512 $\times$ 384, 384 $\times$ 288, 320 $\times$ 240) according to the list of Netflix certified devices~\cite{de2016complexity}.
  \item H.264/HEVC/VP9 compression: We encoded the downsampled sequences using three commonly used video encoders with two-pass encoding~\cite{grois2013performance,de2016complexity,kreuzberger2016comparative}. The target bitrate ranges from 100 kbps to 9 Mbps with a step size of 100 kbps.
\end{itemize}
In total, we obtain 540 (hypothetical reference circuit) $\times$ 250 (source) $\times$ 3 (encoder) = 405,000 test samples (currently the largest in the VQA community). We evaluate the quality of each video at a given spatial resolution, bitrate, and five commonly used display devices including cellphone, tablet, laptop, desktop, and TV using SSIMplus~\cite{Rehman2015SSIMplus} for the following reasons. First, SSIMplus is currently the only HVS motivated spatial resolution and display device-adapted VQA model that is shown to outperform other state-of-the-art quality measures in terms of accuracy and speed~\cite{Rehman2015SSIMplus,Duanmu2017QoE}. Second, a simplified VQA model SSIM~\cite{wang2004image} has been demonstrated to perform well in estimating the GRD functions~\cite{chen2016subjective}. The resulting dense samples of SSIMplus are regarded as the ground truth of GRD functions (The range of SSIMplus is from 0 to 100 with 100 indicating perfect quality). Our GRD modeling approach does not constrain itself on any specific VQA methods. When other ways of generating dense ground-truth samples are available, the same GRD modeling approach may also be applied.

\textbf{Implementation Details:} We initialize the scattered network with delaunay triangulation~\cite{delaunay1934sphere}, inherited from CT method~\cite{clough1965ct}. The balance weight $\lambda$ in~\eqref{eq:robust} is set to $10^{-4}$. In our current experiments, the performance of the proposed RAMCT is fairly insensitive to variations of the value. We employed OSQP~\cite{stellato2017osqp} to solve the quadratic programming problem, where the maximum number of iterations is set to $10^6$. The stopping criteria threshold $T$ is set to 540 (the total number of representation samples in the discretized GRD function space) $\times$ 10 (the standard deviation of mean opinion score in the LIVE Video Quality Assessment database), resulting in an average sample number of 38. When $tr(\Sigma)$ is below the threshold, we conclude that the uncertainty in the system can be explained by the disagreement between subjects. Therefore, further improvement in prediction accuracy may not be as meaningful. Since a triangulation only covers the convex hull of the scattered point set, extrapolation beyond the convex hull is not possible. In order to make a fair comparison, we initialize the training set $S$ as the representations with maximum and minimum bitrates at all spatial resolutions. To construct the covariance matrix described in Section~\ref{sec:sampling} as well as test the proposed algorithm, we randomly segregated the database into a training set of 200 GRD functions and a testing set with 50 GRD functions. The random split is repeated 50 times and the median performance is reported.

\textbf{Evaluation Criteria:} We test the performance of the GRD estimators in terms of both accuracy and rate of convergence. Specifically, we used two metrics to evaluate the accuracy. The mean squared error (MSE) and $l_{\infty}$ norm of the error values are computed between the estimated function and the actual function for each source content. The median results are then computed over all testing functions. All interpolation models can fit increasingly complex GRD functions at the cost of using many parameters. What distinguishes these models from each other is the rate and manner with which the quality of the approximation varies with the number of training samples.

\subsection{Performance}\label{subsec:existing}
We test five GRD function models including reciprocal regression~\cite{toni2015optimal}, logarithmic regression~\cite{chen2016subjective}, 1D piecewise cubic Hermite interpolating polynomial (PCHIP), CT interpolation, and the proposed RAMCT on the aforementioned database. To evaluate the performance of the uncertainty sampling algorithm, we apply it on the five GRD models above and compare its performance with random sampling scheme as the baseline. For random sampling, the initial set of training sample $S$ is set as the representations with the maximum and minimum bitrates at all spatial resolutions to allow fair comparison. The training process with random sampling was repeated 50 times and the median performance is reported.

\begin{figure*}[t]
    \centering
    \captionsetup{justification=centering}
    \subfloat[]{\includegraphics[width=0.33\textwidth]{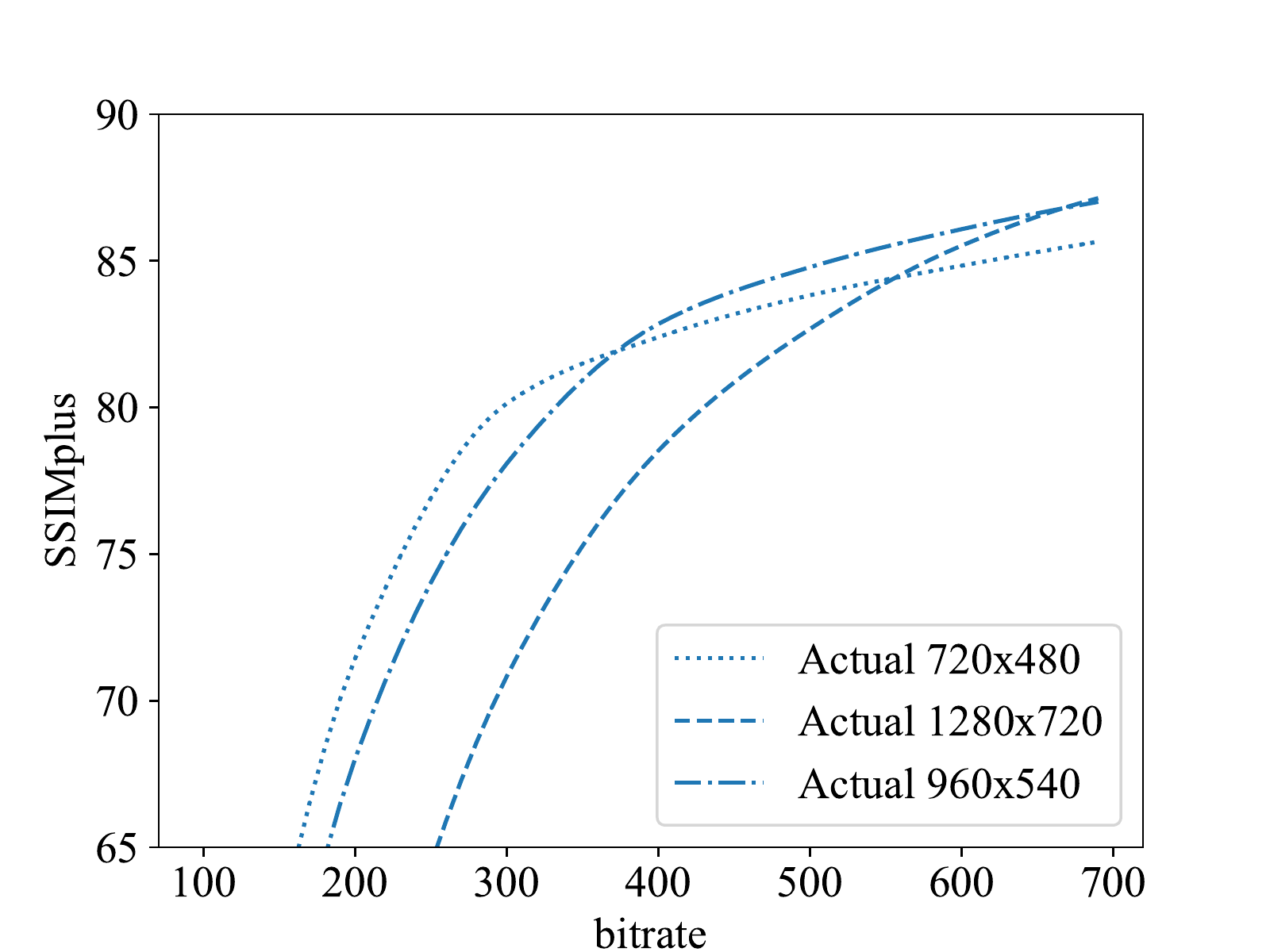}}\hskip.2em
    \subfloat[]{\includegraphics[width=0.33\textwidth]{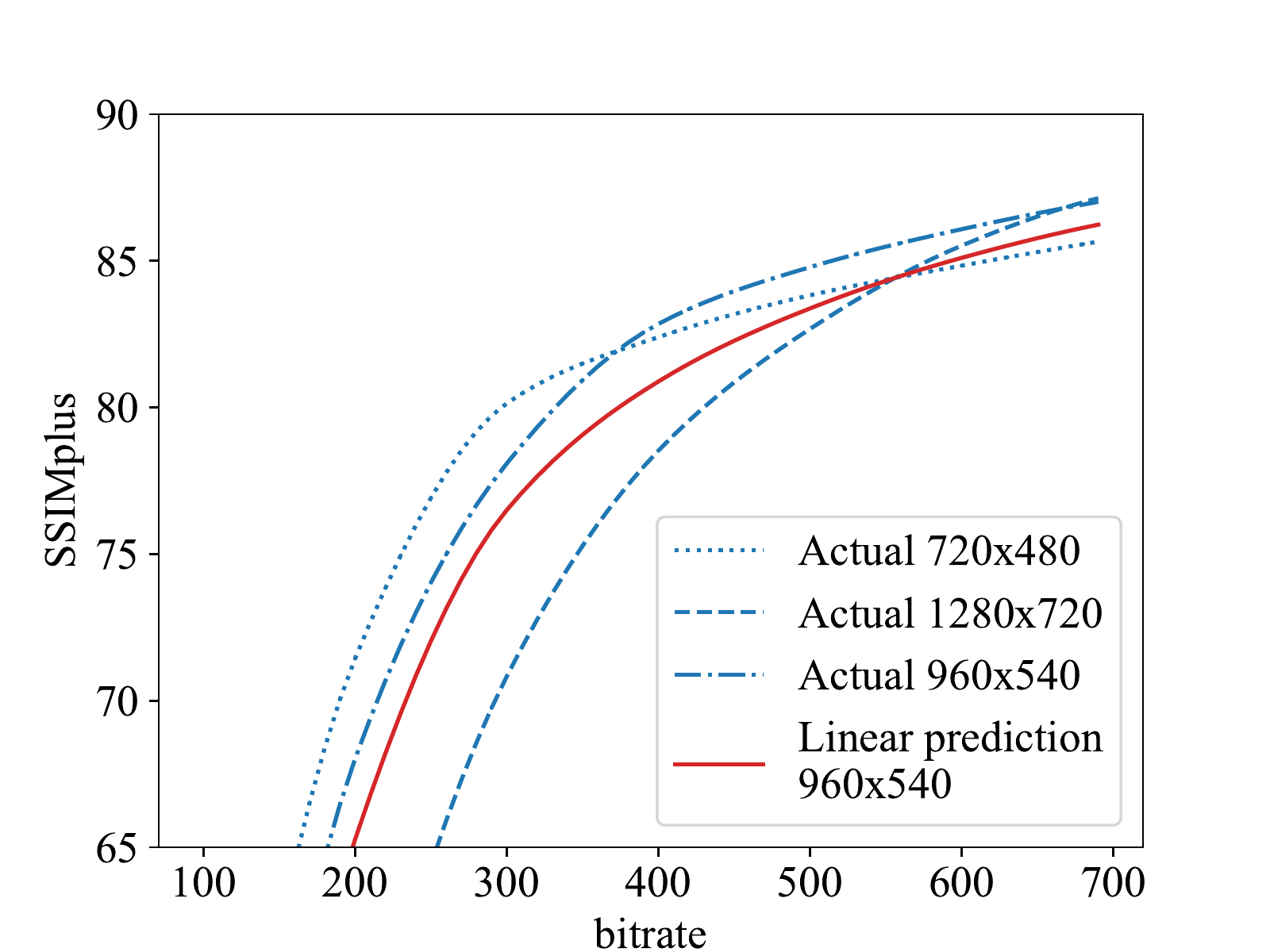}}\hskip.2em
    \subfloat[]{\includegraphics[width=0.33\textwidth]{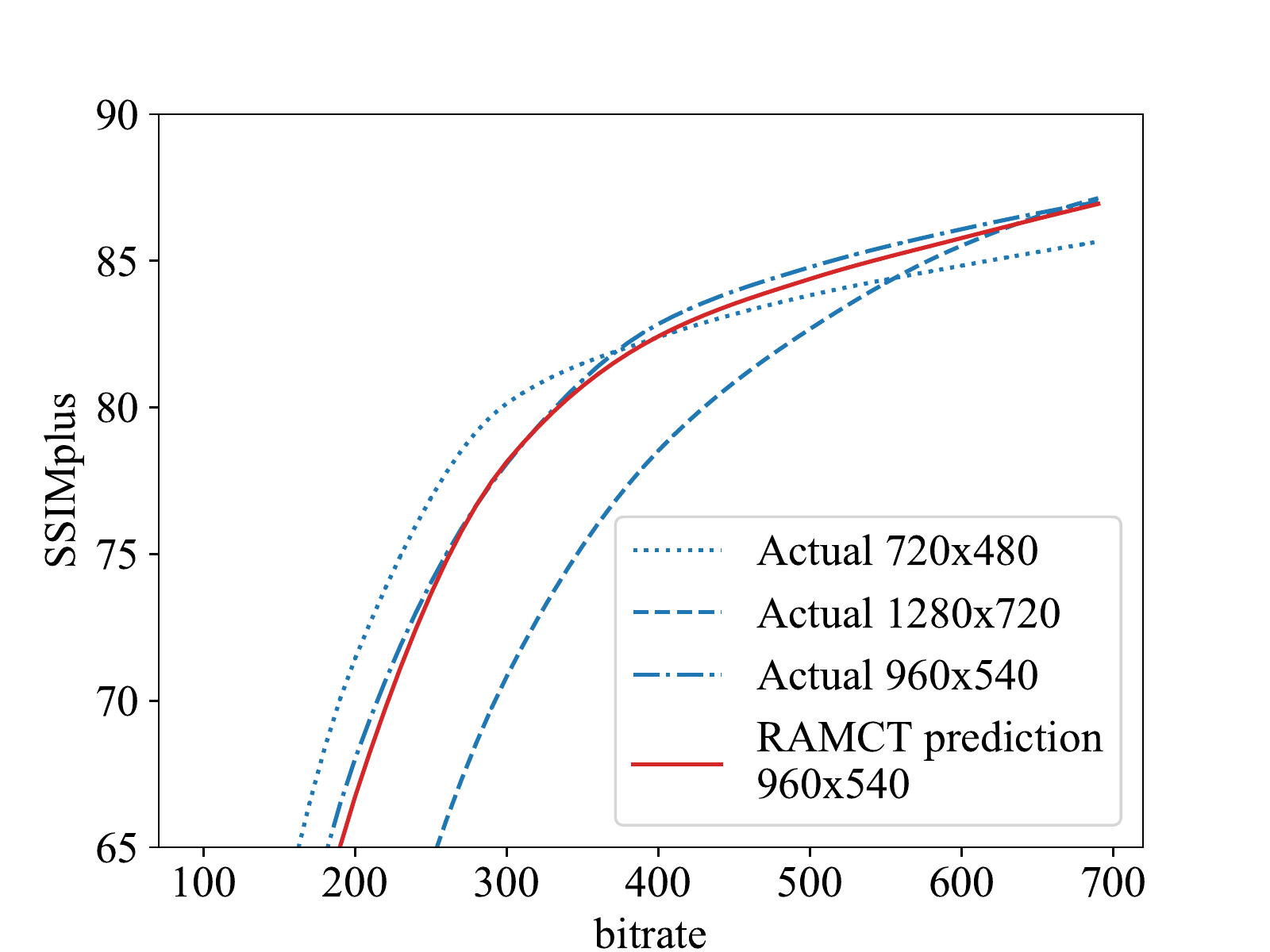}}
    \caption{Prediction of RD curve of novel resolution from known RD curves of other resolutions. (a) Ground truth RD curves; (b) Prediction of 960$\times$540 RD curve from 720$\times$480 and 1280$\times$720 curves using linear interpolation; (c) Prediction of 960$\times$540 RD curve using the proposed method.}\label{fig:rd_curves}
\end{figure*}

Table~\ref{tab:mse} and~\ref{tab:linf} show the prediction accuracy on the database, from which the key observations are summarized as follows. First, the models that assume a certain analytic functional form are consistently biased, failing to accurately fit GRD functions even with all samples probed. On the other hand, the existing interpolation models usually take more than 100 random samples to converge, although they are asymptotically unbiased. By contrast, the proposed RAMCT model converges with only a moderate number of samples. Second, we analyze the core contributors of RAMCT with deliberate selection of competing models. Specifically, the 1D monotonic interpolant PCHIP significantly outperforms the 2D generic interpolant CT, suggesting the importance of axial monotonicity. RAMCT achieves even better performance by exploiting the 2D structure and jointly modeling the GRD functions. Third, we observe strong generalizability of the proposed uncertainty sampling strategy evident by the significant improvement over random sampling on all models. The performance improvement is most salient on the proposed model. In general, RAMCT is able to accurately model GRD functions with only 30 labeled samples, based on which the reciprocal model merely have sufficient known variables to initialize fitting. To gain a concrete impression, we also recorded the execution time of the entire GRD estimation pipeline including video encoding, objective VQA, and GRD function approximation with the competing algorithms on a computer with 3.6GHz CPU and 16G RAM. RAMCT with uncertainty sampling takes around 10 minutes to reduce $l_{\infty}$ below 5, which is more than 100 times faster than the tradition regression models with random sampling.

\section{Applications}
The application scope of GRD model is much broader than VQA. Here we demonstrate three use cases of the proposed model.

\subsection{Rate-Distortion Curve at Novel Resolutions}
Given a set of rate-distortion curves at multiple resolutions, it is desirable to predict the rate-distortion performance at novel resolutions, especially when there exists a mismatch between the supported viewing device of downstream content delivery network and the recommended encoding profiles. Traditional methods linearly interpolate the rate-distortion curve at novel resolutions~\cite{de2016complexity}, neglecting the characteristics of GRD functions. Fig.~\ref{fig:rd_curves} compares the linearly interpolated and RAMCT-interpolated rate-distortion curves at 960$\times$540 with the ground truth SSIMplus curve, from which we have several observations. First, the linearly interpolated curve shares the same intersection with the neighboring curves at 740$\times$480 and 1280$\times$720, inducing consistent bias to the prediction. The proposed RAMCT model is able to accurately predict the quality at the intersection of the neighboring curves by taking all known rate-distortion curves into consideration. Second, the linearly interpolated rate-distortion curve always lies between its neighboring curves, suggesting that the predicted quality at any bitrate is lower than the quality on one of its neighboring curves. This behavior contradicts the fact that each resolution may have a bitrate region in which it outperforms other resolutions~\cite{de2016complexity}. On the contrary, RAMCT better preserves the general trend of resolution-quality curve at different bitrate, thanks to the regularization imposed by the $C^1$ condition at given nodes. Third, RAMCT outperforms the linear interpolation model in predicting the ground truth rate-distortion curve across all bitrates. The experimental results also justify the effectiveness of the $C^1$ and smoothness prior used in RAMCT.


\begin{table}
  \centering
  \caption{$l_{\infty}$ performance of the competing GRD function models with different number of labeled samples selected by random sampling (RS) and the proposed uncertainty sampling (US).}\label{tab:rd_curve_results}
  \begin{tabular}{c|C{1cm}C{1cm}|C{1cm}C{1cm}}
      \toprule
     \multirow{2}{*}{Resolution}   & \multicolumn{2}{c|}{$l_{\infty}$} & \multicolumn{2}{c}{MSE} \\ \cline{2-5}
                                   & Linear & RAMCT & Linear & RAMCT \\ \hline
     640$\times$360                & 7.68  & 3.12  & 7.89  & 1.56 \\
     960$\times$540                & 7.66  & 4.83  & 6.61  & 3.14  \\
     1600$\times$900               & 8.77  & 7.87  & 5.18  & 4.98 \\ \hline
     Average                       & 8.04  & 5.27  & 6.56  & 3.23 \\
     \bottomrule
   \end{tabular}
\end{table}

To further validate the performance of the proposed GRD model at novel spatial resolutions, we predict the rate-distortion curves of 20 randomly selected source videos from the dataset at three novel resolutions (640$\times$360, 960$\times$540, and 1600$\times$900). The evaluated bitrate ranges from 100 kbps to 9 Mbps with a step size of 100 kbps. The results are listed in Table~\ref{tab:rd_curve_results}. We can observe that RAMCT outperforms the linear model~\cite{de2016complexity} with a clear margin at novel resolutions.

\subsection{Per-Title Encoding Profile Generation}
\begin{figure*}[t]
    \centering
    \captionsetup{justification=centering}
    \subfloat[Title with low complexity]{\includegraphics[width=0.33\textwidth]{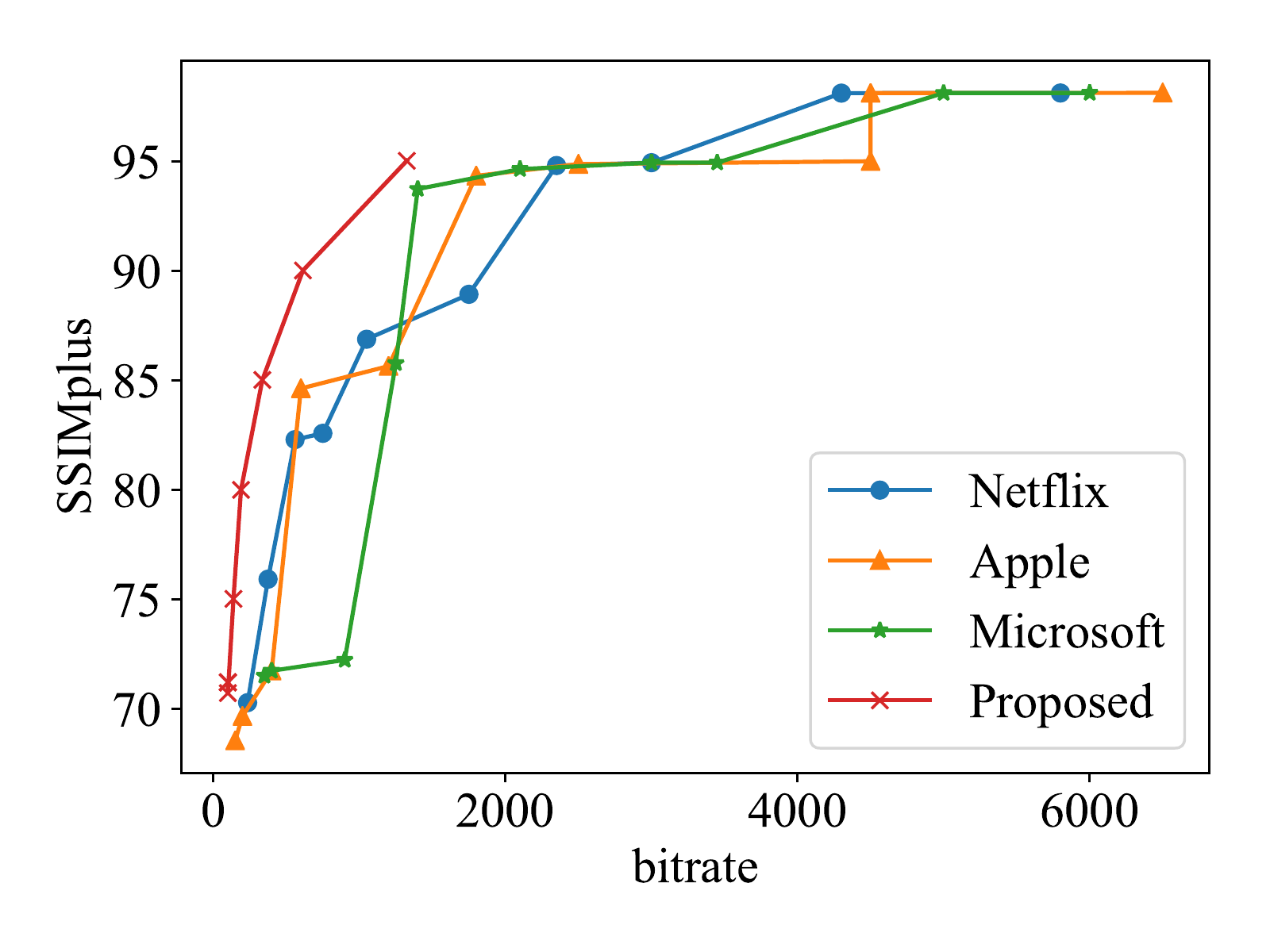}}\hskip.2em
    \subfloat[Title with moderate complexity]{\includegraphics[width=0.33\textwidth]{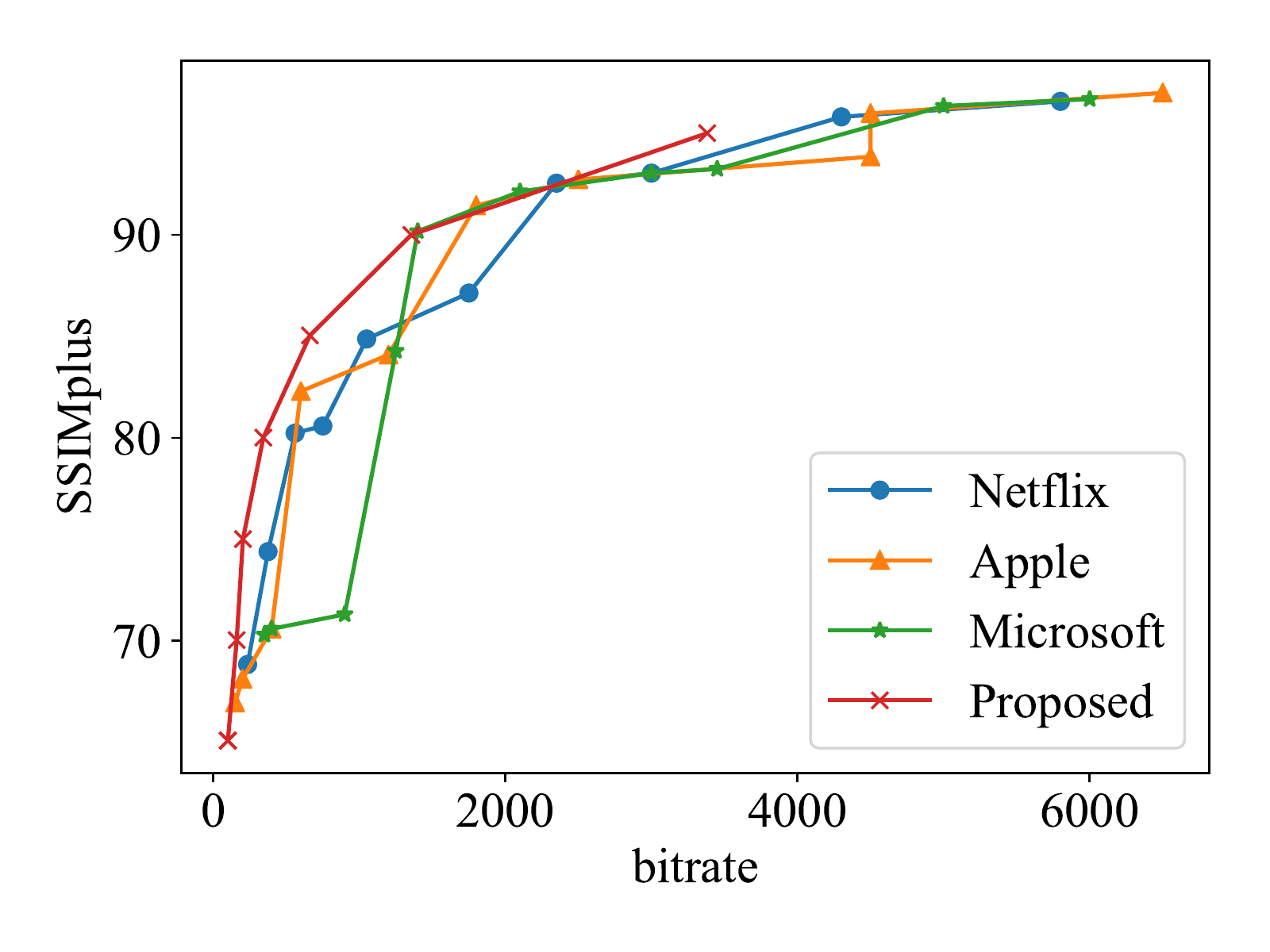}}\hskip.2em
    \subfloat[Title with high complexity]{\includegraphics[width=0.33\textwidth]{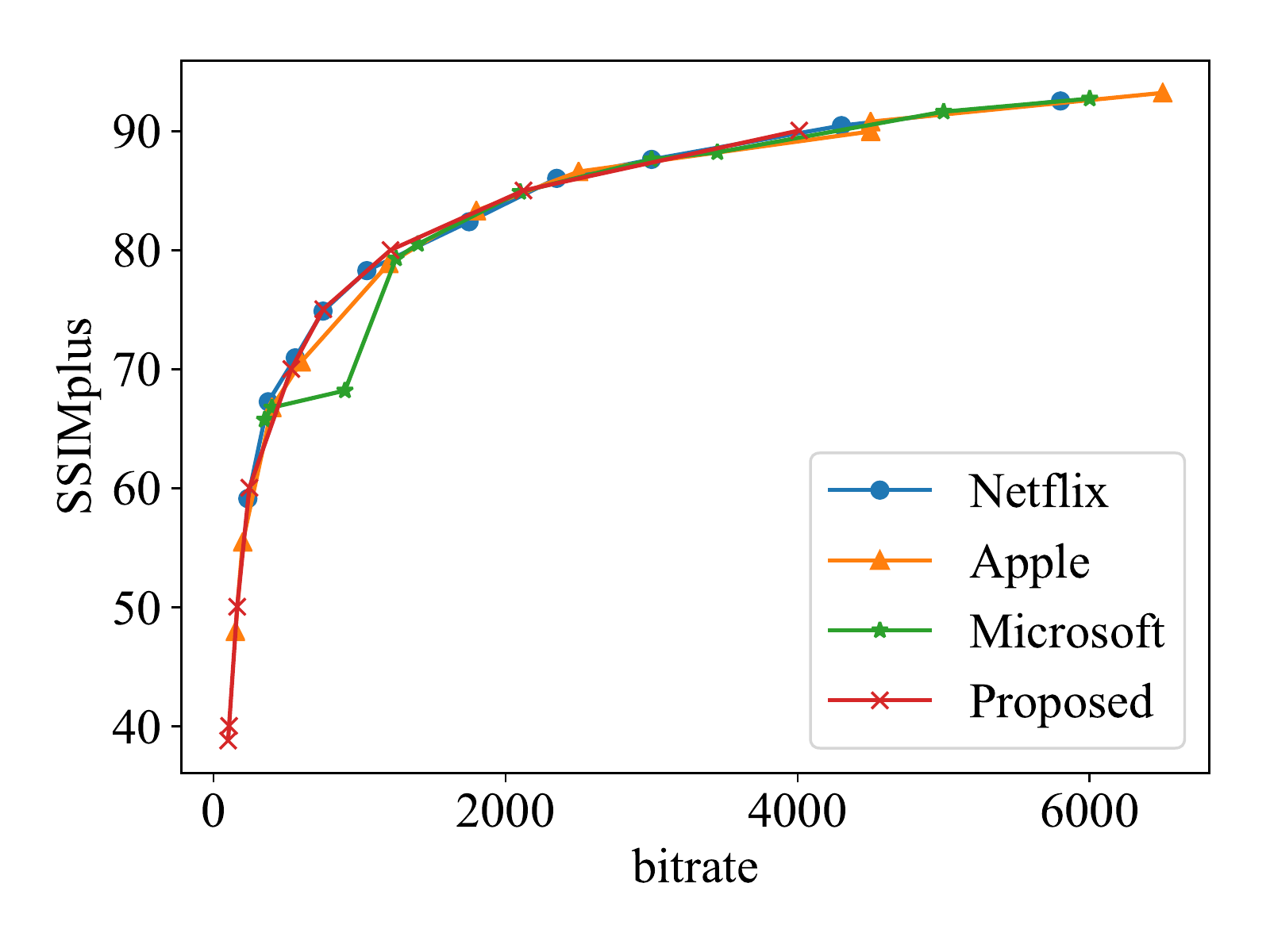}}
    \caption{Bitrate ladders generated by the recommendations and the proposed algorithm for three contents.}\label{fig:pertitle}
\end{figure*}

To overcome the heterogeneity in users' network conditions and display devices, video service providers often encode videos at multiple bitrates and spatial resolutions. However, the selection of the encoding profiles are either hard-coded, resulting in sub-optimal QoE due to the negligence of the difference in source video complexities, or selected based on interactive objective measurement and subjective judgement that are inconsistent and time-consuming. To deliver the best quality video to consumers, each title should receive a unique bitrate ladder, tailored to its specific complexity characteristics. We introduce a quality-driven per-title optimization framework to automatically select the best encoding configurations, where the proposed GRD model serves as the key component.

Content delivery networks often aim to deliver videos at certain quality levels to satisfy different viewers. It is beneficial to minimize the bitrate usage in the encoding profile when achieving the objective. Mathematically, the quality-driven bitrate ladder selection problem can be formulated as a constrained optimization problem. Specifically, for the $i$-th representation,
\begin{equation*}
\begin{aligned}
& \underset{ \{x, y\} }{\text{minimize}}
& & x \\
& \text{subject to}
& & f(x, y) \geq C_i, \; i = 1, \ldots, m,
\end{aligned}
\end{equation*}
where $x$, $y$, $f(\cdot, \cdot)$, $C_i$ and $m$ represent the bitrate, the spatial resolution, the GRD function, the target quality level of video representation $i$, and the total number of video representations, respectively. Solving the optimization problem requires precise knowledge of the GRD function. Thanks to the effectiveness and differentiability of RAMCT, the proposed model can be incorporated with gradient-based optimization tools~\cite{cvx} to solve the per-title optimization problem. (Interested readers may refer to the appendix for more details on how we solve the optimization problem).

\begin{table}
\centering
\caption{Average bitrate saving of encoding profiles. Negative values indicate actual bitrate reduction}
\label{tab:pertitle_bdrate}
\begin{tabular}{c|cccc}
\toprule
              &   Microsoft   &   Apple   &   Netflix   &   Proposed \\ \hline
Microsoft     &       0       &     -     &    -        &    -    \\
Apple         &   -25.3\%     &     0     &    -        &    -    \\
Netflix       &   -29.3\%     &  -5.6\%   &    0        &    -    \\
Proposed      &   -62.0\%     & -48.9\%   & -46.8\%     &    0    \\
\bottomrule
\end{tabular}
\end{table}

To validate the proposed per-title encoding profile selection algorithm, we apply the algorithm to generate bitrate ladders using H.264~\cite{wiegand2003overview} for 50 randomly selected videos in the aforementioned dataset. We set the target quality levels $\{C_i\}_{i=1}^{10}$ as $\{30, 40, 50, 60, 70, 75, 80, 85, 90, 95\}$ to cover diverse quality range and to match the total number of representations in standard recommendations~\cite{toni2015optimal}. For simplicity, we optimize the representation sets for only one viewing device (cellphone), while the procedure can be readily extended to multiple devices to generate a more comprehensive representation set. In Fig.~\ref{fig:pertitle}, we compare the rate-quality curve of representation sets generated by the proposed algorithm, recommendations by Netflix~\cite{netflix2015pertitle}, Apple~\cite{applea}, and Microsoft~\cite{grafl2013combined} for three videos with different complexities, from which the key observations are as follows. First, contrasting the hand-crafted bitrate ladders, the encoding profile generated by the proposed algorithm is content adaptive. Specifically, the encoding bitrate increases with respect to the complexity of the source video as illustrated from Fig.~\ref{fig:pertitle}(a) to Fig.~\ref{fig:pertitle}(c). Second, the proposed method achieves the highest quality at all bitrate levels. The performance improvement is mainly introduced by the encoding strategy at the convex hull encompassing the individual per-resolution rate-distortion curves~\cite{de2016complexity}. Table~\ref{tab:pertitle_bdrate} provides a full summary of the Bj{\o}ntegaard-Delta bitrate (BD-Rate)~\cite{bjontegaard2001bda}, indicating the required overhead in bitrate to achieve the same SSIMplus values. We observe that the proposed framework outperforms the existing hard-coded representation sets by at least 47\%.
 
\subsection{Codec Comparison}
\begin{figure}
  \centering
  \includegraphics[width=0.35\textwidth]{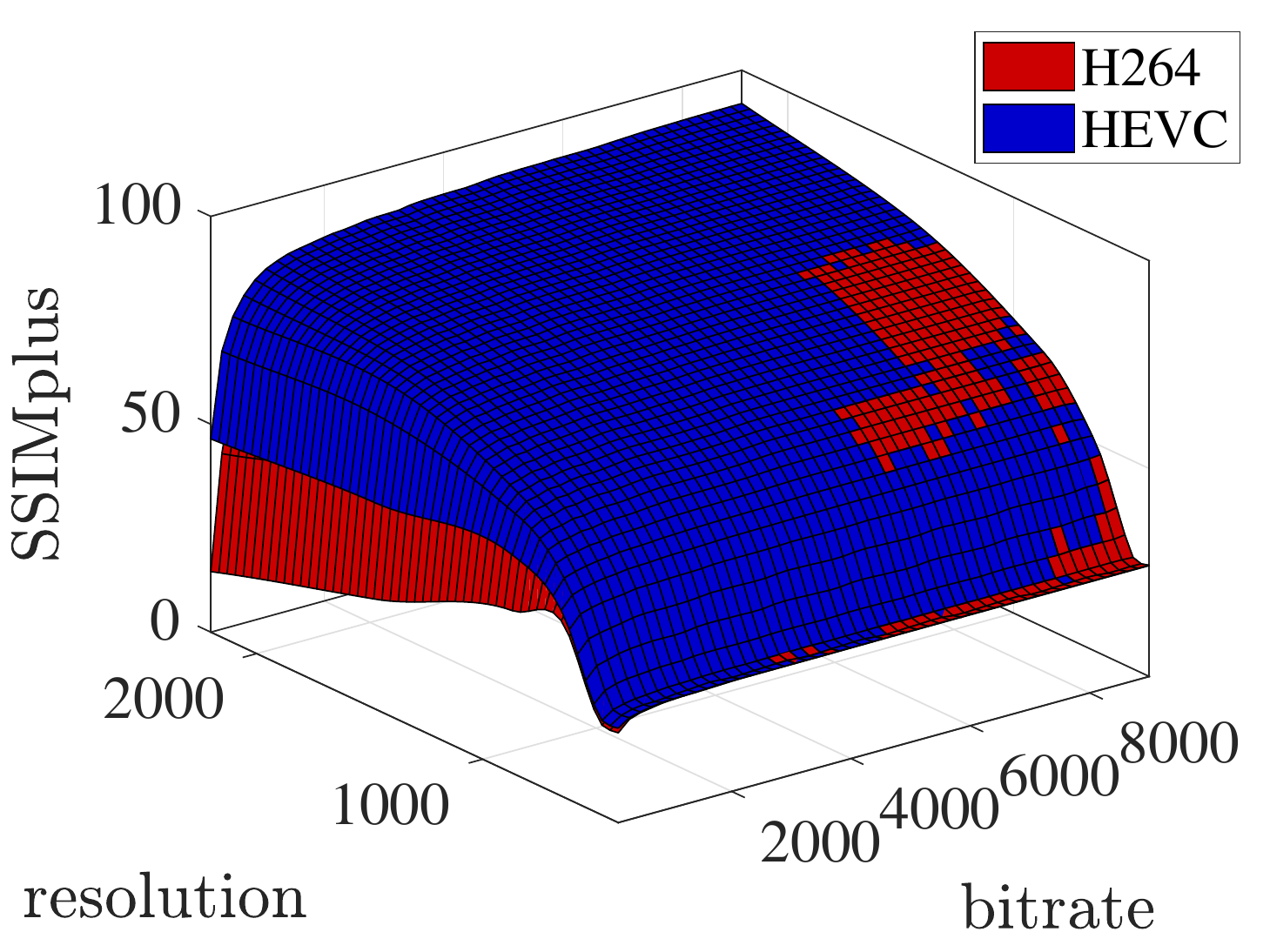}
  \caption{Generalized rate-distortion surfaces of H.264 and HEVC encoders for a sample source video.}\label{fig:codec_samples}
\end{figure}

In the past decade, there has been a tremendous growth in video compression algorithms, thanks to the fast development of computational multimedia. With many video encoders at hand, it becomes pivotal to compare their performance, so as to find the best algorithm as well as directions for further advancement. Bj{\o}ntegaard-Delta model~\cite{bjontegaard2001bda,bjontegaard2008bdb} has become the most commonly used objective coding efficiency measurement. Bj{\o}øntegaard-Delta PSNR (BD-PSNR) and BD-Rate are typically computed as the difference in bitrate and quality (measured in PSNR) based on the interpolated rate-distortion curves
\begin{subequations}\label{eq:bd}
  \begin{align}
  Q_{BD}=& \frac{\int_{x_L}^{x_H} [z_B(x) - z_A(x)] dx}{\int_{x_L}^{x_H}dx}, \\
  R_{BD}\approx& 10^{\frac{\int_{z_L}^{z_H}[x_B(z)-x_A(z)]dz}{\int_{z_L}^{z_H} dz}} - 1,
  \end{align}
\end{subequations}
where $x_A$ and $x_B$ are the logarithmic-scale bitrate, $z_A$ and $z_B$ are the quality of the interpolated reference and test bitrate curves, respectively. $[x_L, x_H]$ and $[z_L, z_H]$ are the effective domain and range of the rate-distortion curves. However, BD-PSNR and BD-Rate do not take spatial resolution and viewing condition into consideration. Fig.~\ref{fig:codec_samples} shows two GRD surfaces generated by H.264~\cite{wiegand2003overview} and HEVC encoders for a source video. Although H.264 performs on par with HEVC at low resolutions, it requires higher bitrate to achieve the same target quality at high resolutions. Therefore, applying BD-Rate on a single resolution is not sufficient to fairly compare the overall performance between encoders. To this regard, we propose generalized quality gain ($Q_{gain}$) and rate gain ($R_{gain}$) models as
\begin{subequations}\label{eq:gbd}
  \begin{align}
  Q_{gain}=& \frac{\int_{U}\int_{y_L}^{y_H}\int_{x_L}^{x_H} p(u)[z_B(x, y, u) - z_A(x, y, u)] dxdydu}{\int_{y_L}^{y_H}\int_{x_L}^{x_H}dxdy}, \\
  R_{gain}\approx& 10^{\frac{\int_U\int_{y_L}^{y_H}\int_{z_L}^{z_H}p(u)[x_B(z, y, u)-x_A(z, y, u)]dzdydu}{\int_U\int_{y_L}^{y_H}\int_{z_L}^{z_H}p(u)dzdydu}} - 1,
  \end{align}
\end{subequations}
where $p(u)$, $U$, and $[y_L, y_H]$ represent the probability density of viewing devices, the set of all device of interests, and the domain of video spatial resolution, respectively. The generalized $Q_{gain}$ and $R_{gain}$ models represent the expected quality gain and the expected bitrate gain (saving when $R_{gain}$ negative) across all spatial resolutions and viewing devices, leading to a more comprehensive evaluation of codecs. It should be noted that $z_A(x, y, u)$ is essentially the GRD function of codec $A$, which can be efficiently approximated by the proposed model. $x_A(z, y, u)$ can also be estimated numerically from the interpolated surface. (Interested readers may refer to the appendix for more details on how we compute $Q_{gain}$ and $R_{gain}$.) Therefore, RAMCT is a natural fit to the generalized $Q_{gain}$ and $R_{gain}$ models. The effect of any individual influencing factor can be obtained by taking the marginal expectation in the corresponding dimension, which is more robust than BD-PSNR and BD-Rate at a single resolution.

\begin{table}
  \centering
  \caption{Performance of VP9, and HEVC in terms of the generalized $Q_{gain}$ and $R_{gain}$ models with H.264 as the baseline codec}\label{tab:gbd_result}
  \begin{tabular}{c|C{1cm}C{1cm}}
      \toprule
     Codec   & $Q_{gain}$ & $R_{gain}$ \\ \hline
     VP9     & 1.9 & -27.5\% \\
     HEVC    & 0.2 & -15.9\%  \\
     \bottomrule
   \end{tabular}
\end{table}

Using the proposed measures on the aforementioned dataset, we evaluate the performance of three video encoders, namely H.264, VP9, and HEVC. In order to simplify the expression, we set $p(u)$ to be uniform distribution for five display devices including cellphone, tablet, laptop, desktop, and TV in the paper. Table~\ref{tab:gbd_result} shows the performance of VP9 and HEVC in terms of the proposed measures with H.264 as the reference codec, from which we can observe that VP9 outperforms HEVC by an average of 1.7 and 11\% in $Q_{gain}$ and $R_{gain}$, respectively. The results of our objective codec evaluation are in general consistent with the recent video quality assessment results~\cite{msu2017codec,bitmovin2018codec}.

\section{Conclusions}\label{sec:conclusion}
GRD functions represent the critical link between multimedia resource and perceptual QoE. In this work, we proposed a learning framework to model the GRD function by exploiting the properties all GRD functions share and the information redundancy of training samples. The framework leads to an efficient algorithm that demonstrates state-of-the-art performance, which we believe arises from the RAMCT model for imposing axial monotonicity, the joint modeling of the multi-dimensional GRD function for exploiting its functional structure, and the information-theoretic sampling algorithm for improving the quality of training samples. Extensive experiments have shown that the algorithm is able to accurately model the function with a very small number of training samples. Furthermore, we demonstrate that the proposed GRD model plays a central role in a great variety of visual communication applications.

The current work can be extended in many ways. As a basis for future work, we note that the interpolant can be readily extended to higher dimensions~\cite{farin1985modified}, making it applicable to more general applications. For example, in the fields of machine learning~\cite{bishop2006pattern} and data visualization~\cite{sarfraz2006visualize}, flexible monotonic interpolation can provide regularization and makes the model more interpretable. Another promising future direction is to develop models that can predict GRD functions without sampling the GRD space.

\bibliographystyle{IEEEtran}
\bibliography{ref} 

\begin{thebibliography}{10}
\providecommand{\url}[1]{#1}
\csname url@samestyle\endcsname
\providecommand{\newblock}{\relax}
\providecommand{\bibinfo}[2]{#2}
\providecommand{\BIBentrySTDinterwordspacing}{\spaceskip=0pt\relax}
\providecommand{\BIBentryALTinterwordstretchfactor}{4}
\providecommand{\BIBentryALTinterwordspacing}{\spaceskip=\fontdimen2\font plus
\BIBentryALTinterwordstretchfactor\fontdimen3\font minus
  \fontdimen4\font\relax}
\providecommand{\BIBforeignlanguage}[2]{{%
\expandafter\ifx\csname l@#1\endcsname\relax
\typeout{** WARNING: IEEEtran.bst: No hyphenation pattern has been}%
\typeout{** loaded for the language `#1'. Using the pattern for}%
\typeout{** the default language instead.}%
\else
\language=\csname l@#1\endcsname
\fi
#2}}
\providecommand{\BIBdecl}{\relax}
\BIBdecl

\bibitem{grois2013performance}
D.~Grois, D.~Marpe, A.~Mulayoff, B.~Itzhaky, and O.~Hadar, ``Performance
  comparison of {H.265/MPEG-HEVC}, {VP9}, and {H.264/MPEG-AVC} encoders,'' in
  \emph{Picture Coding Symposium}, 2013, pp. 394--397.

\bibitem{wang2012ssim}
S.~Wang, A.~Rehman, Z.~Wang, S.~Ma, and W.~Gao, ``{SSIM}-motivated
  rate-distortion optimization for video coding,'' \emph{IEEE Trans. Circuits
  and Systems for Video Tech.}, vol.~22, no.~4, pp. 516--529, Apr. 2012.

\bibitem{ou2014q}
Y.~Ou, Y.~Xue, and Y.~Wang, ``{Q-STAR}: {A} perceptual video quality model
  considering impact of spatial, temporal, and amplitude resolutions,''
  \emph{IEEE Trans. Image Processing}, vol.~23, no.~6, pp. 2473--2486, Jun.
  2014.

\bibitem{zhang2013qoe}
W.~Zhang, Y.~Wen, Z.~Chen, and A.~Khisti, ``{QoE}-driven cache management for
  {HTTP} adaptive bit rate streaming over wireless networks,'' \emph{IEEE
  Trans. Multimedia}, vol.~15, no.~6, pp. 1431--1445, Oct. 2013.

\bibitem{toni2015optimal}
L.~Toni, R.~Aparicio-Pardo, K.~Pires, G.~Simon, A.~Blanc, and P.~Frossard,
  ``Optimal selection of adaptive streaming representations,'' \emph{ACM Trans.
  Multimedia Computing, Communications, and Applications}, vol.~11, no.~2, pp.
  1--43, Feb. 2015.

\bibitem{de2016complexity}
J.~De~Cock, Z.~Li, M.~Manohara, and A.~Aaron, ``Complexity-based
  consistent-quality encoding in the cloud,'' in \emph{Proc. IEEE Int. Conf.
  Image Proc.}, 2016, pp. 1484--1488.

\bibitem{chen2016subjective}
C.~Chen, S.~Inguva, A.~Rankin, and A.~Kokaram, ``A subjective study for the
  design of multi-resolution {ABR} video streams with the {VP9} codec,'' in
  \emph{Electronic Imaging}, 2016, pp. 1--5.

\bibitem{wang2015objective}
Z.~Wang, K.~Zeng, A.~Rehman, H.~Yeganeh, and S.~Wang, ``Objective video
  presentation {QoE} predictor for smart adaptive video streaming,'' in
  \emph{Proc. SPIE Optical Engineering+Applications}, 2015, pp.
  95\,990Y.1--95\,990Y.13.

\bibitem{chen2017encoding}
C.~Chen, Y.~Lin, A.~Kokaram, and S.~Benting, ``Encoding bitrate optimization
  using playback statistics for {HTTP}-based adaptive video streaming,''
  \emph{arXiv preprint arXiv:1709.08763}, Sep. 2017.

\bibitem{ISO2012Dash}
\BIBentryALTinterwordspacing
D.~I. Forum. (2013) For promotion of {MPEG-DASH} 2013. [Online]. Available:
  \url{http://dashif.org.}
\BIBentrySTDinterwordspacing

\bibitem{aom2018AV1}
\BIBentryALTinterwordspacing
{Alliance for Open Media}. (2018) Av1 bitstream and decoding process
  specification. [Online]. Available:
  \url{https://aomedia.org/av1-bitstream-and-decoding-process-specification/.}
\BIBentrySTDinterwordspacing

\bibitem{li2016VMAF}
\BIBentryALTinterwordspacing
Z.~Li, A.~Aaron, L.~Katsavounidis, A.~Moorthy, and M.~Manohara. (2016) Toward a
  practical perceptual video quality metric. [Online]. Available:
  \url{http://techblog.netflix.com/2016/06/toward-practical-perceptual-video.html.}
\BIBentrySTDinterwordspacing

\bibitem{kreuzberger2016comparative}
C.~Kreuzberger, B.~Rainer, H.~Hellwagner, L.~Toni, and P.~Frossard, ``A
  comparative study of {DASH} representation sets using real user
  characteristics,'' in \emph{Proc. Int. Workshop on Network and OS Support for
  Digital Audio and Video}, 2016, pp. 1--4.

\bibitem{berger1975rate}
T.~Berger, ``Rate distortion theory and data compression,'' in \emph{Advances
  in Source Coding}.\hskip 1em plus 0.5em minus 0.4em\relax Springer, 1975, pp.
  1--39.

\bibitem{zhai2008cross}
G.~Zhai, J.~Cai, W.~Lin, X.~Yang, W.~Zhang, and M.~Etoh, ``Cross-dimensional
  perceptual quality assessment for low bit-rate videos,'' \emph{IEEE Trans.
  Multimedia}, vol.~10, no.~7, pp. 1316--1324, Nov. 2008.

\bibitem{robson1966spatial}
J.~Robson, ``Spatial and temporal contrast-sensitivity functions of the visual
  system,'' \emph{Journal of Optical Society of America}, vol.~56, no.~8, pp.
  1141--1142, Aug. 1966.

\bibitem{clough1965ct}
R.~Clough and T.~J., ``Finite element stiffness matrices for analysis of plates
  in bending,'' in \emph{Proceedings of Conf. on Matrix Methods in Structural
  Analysis}, 1965.

\bibitem{alfeld1984trivariate}
P.~Alfeld, ``A trivariate {C}lough-{T}ocher scheme for tetrahedral data,''
  \emph{Computer Aided Geometric Design}, vol.~1, no.~2, pp. 169--181, Jun.
  1984.

\bibitem{amidror2002scattered}
I.~Amidror, ``Scattered data interpolation methods for electronic imaging
  systems: {A} survey,'' \emph{Journal of Electronic Imaging}, vol.~11, no.~2,
  pp. 157--177, Apr. 2002.

\bibitem{farin1980bezier}
G.~Farin, ``B{\'e}zier polynomials over triangles and the construction of
  piecewise {C$^R$} polynomials,'' \emph{Brunel University Mathematics
  Technical Papers collection}, 1980.

\bibitem{nielson1983method}
G.~M. Nielson, ``A method for interpolating scattered data based upon a minimum
  norm network,'' \emph{Mathematics of Computation}, vol.~40, no. 161, pp.
  253--271, 1983.

\bibitem{farin1985modified}
G.~Farin, ``A modified {C}lough-{T}ocher interpolant,'' \emph{Computer Aided
  Geometric Design}, vol.~2, no. 1-3, pp. 19--27, 1985.

\bibitem{fritsch1980monotone}
N.~Fritsch and R.~Carlson, ``Monotone piecewise cubic interpolation,''
  \emph{SIAM Journal on Numerical Analysis}, vol.~17, no.~2, pp. 238--246,
  1980.

\bibitem{han1997fitting}
L.~Han and L.~Schumaker, ``Fitting monotone surfaces to scattered data using c1
  piecewise cubics,'' \emph{SIAM Journal on Numerical Analysis}, vol.~34,
  no.~2, pp. 569--585, Apr. 1997.

\bibitem{stellato2017osqp}
B.~Stellato, G.~Banjac, P.~Goulart, A.~Bemporad, and S.~Boyd, ``{OSQP}: {A}n
  operator splitting solver for quadratic programs,'' \emph{ArXiv preprint
  arXiv:1711.08013}, Nov. 2017.

\bibitem{cortes1995svm}
C.~Cortes and V.~Vapnik, ``Support-vector networks,'' \emph{Machine learning},
  vol.~20, no.~3, pp. 273--297, Sep. 1995.

\bibitem{bishop2006pattern}
C.~Bishop, \emph{Pattern Recognition and Machine Learning}.\hskip 1em plus
  0.5em minus 0.4em\relax Berlin, Heidelberg: Springer-Verlag, 2006.

\bibitem{Rehman2015SSIMplus}
A.~Rehman, K.~Zeng, and Z.~Wang, ``Display device-adapted video
  {Q}uality-of-{E}xperience assessment,'' in \emph{Proc. SPIE}, 2015, pp.
  939\,406.1--939\,406.11.

\bibitem{Duanmu2017QoE}
Z.~Duanmu, K.~Ma, and Z.~Wang, ``Quality-of-{E}xperience of adaptive video
  streaming: {E}xploring the space of adaptations,'' in \emph{Proc. ACM Int.
  Conf. Multimedia}, 2017, pp. 1752--1760.

\bibitem{wang2004image}
Z.~Wang, A.~Bovik, H.~Sheikh, and E.~Simoncelli, ``Image quality assessment:
  {F}rom error visibility to structural similarity,'' \emph{IEEE Trans. Image
  Processing}, vol.~13, no.~4, pp. 600--612, Apr. 2004.

\bibitem{delaunay1934sphere}
B.~Delaunay, ``Sur la sphere vide,'' \emph{Izv. Akad. Nauk SSSR, Otdelenie
  Matematicheskii i Estestvennyka Nauk}, vol.~7, no. 793-800, pp. 1--2, Oct.
  1934.

\bibitem{cvx}
\BIBentryALTinterwordspacing
M.~Grant and S.~Boyd. (2014) {CVX}: Matlab software for disciplined convex
  programming, version 2.1. [Online]. Available: \url{http://cvxr.com/cvx.}
\BIBentrySTDinterwordspacing

\bibitem{wiegand2003overview}
T.~Wiegand, G.~J. Sullivan, G.~Bjontegaard, and A.~Luthra, ``Overview of the h.
  264/avc video coding standard,'' \emph{IEEE Trans. Circuits and Systems for
  Video Tech.}, vol.~13, no.~7, pp. 560--576, Jul. 2003.

\bibitem{netflix2015pertitle}
\BIBentryALTinterwordspacing
A.~Aaron, Z.~Li, M.~Manohara, D.~J. Cock, and D.~Ronca. (2015) {Per-Title}
  encode optimization. [Online]. Available:
  \url{https://medium.com/netflix-techblog/per-title-encode-optimization-7e99442b62a2.}
\BIBentrySTDinterwordspacing

\bibitem{applea}
\BIBentryALTinterwordspacing
Apple. (2016) Best practices for creating and deploying {HTTP} live streaming
  media for {iPhone} and {iPad}. [Online]. Available:
  \url{http://is.gd/LBOdpz.}
\BIBentrySTDinterwordspacing

\bibitem{grafl2013combined}
\BIBentryALTinterwordspacing
G.~Michael, T.~Christian, H.~Hermann, C.~Wael, N.~Daniel, and B.~Stefano,
  ``Combined bitrate suggestions for multi-rate streaming of industry
  solutions,'' 2013. [Online]. Available:
  \url{http://alicante.itec.aau.at/am1.html.}
\BIBentrySTDinterwordspacing

\bibitem{bjontegaard2001bda}
G.~Bj{\o}ntegaard, ``Calculation of average {PSNR} differences between
  rd-curves,'' Austin, TX, USA, Tech. Rep. VCEG-M33, ITU-T SG 16/Q6, 13th VCEG
  Meeting, Apr. 2001.

\bibitem{bjontegaard2008bdb}
------, ``Improvements of the {BD-PSNR} model,'' Berlin, Germany, Tech. Rep.
  VCEG-AI11, ITU-T SG 16/Q6, 35th VCEG Meeting, Jul. 2008.

\bibitem{msu2017codec}
\BIBentryALTinterwordspacing
D.~Vatolin, D.~Kulikov, E.~Mikhail, D.~Stanislav, and Z.~Sergey. (2017) {MSU}
  codec comparison 2017 part {V}: {High} quality encoders. [Online]. Available:
  \url{http://www.compression.ru/video/codec_comparison/hevc_2017/MSU_HEVC_comparison_2017_P5_HQ_encoders.pdf.}
\BIBentrySTDinterwordspacing

\bibitem{bitmovin2018codec}
\BIBentryALTinterwordspacing
F.~Christian. (2018) Multi-codec {DASH} dataset: {An} evaluatation of {AV1},
  {AVC}, {HEVC} and {VP9}. [Online]. Available:
  \url{https://bitmovin.com/av1-multi-codec-dash-dataset/.}
\BIBentrySTDinterwordspacing

\bibitem{sarfraz2006visualize}
M.~Sarfraz and M.~Hussain, ``Data visualization using rational spline
  interpolation,'' \emph{Journal of Computational and Applied Mathematics},
  vol. 189, no.~1, pp. 513--525, May 2006.

\end{thebibliography}

\end{document}


\title{Robust Axial-Monotonic Clough-Tocher Model for Generalized Rate-Distortion Functions}
\author{Zhengfang~Duanmu,~\IEEEmembership{Student Member,~IEEE,}\\
        Wentao~Liu,~\IEEEmembership{Student Member,~IEEE,}
        and~Zhou~Wang,~\IEEEmembership{Fellow,~IEEE}
\thanks{ The authors are with the Department of Electrical and Computer Engineering, University of Waterloo, Waterloo, ON N2L 3G1, Canada (e-mail: \{zduanmu, w238liu, zhou.wang\}@uwaterloo.ca).}
}

\markboth{}%
{Shell \MakeLowercase{\textit{et al.}}: Bare Demo of IEEEtran.cls for Journals}

\appendices
\section{Conversion from Unknowns to B\'ezier Coordinate}
Summarizing (2) as matrix form, we obtain
\begin{align}
\nonumber {\bf c} = {\bf R}{\bf d}+{\bf f}
\end{align}
More specifically,
\begin{align}
\nonumber {\bf c}&=\left[
\scalemath{0.9}{
\begin{array}{*{20}{c}}
c_{T_{01}} \\ 
c_{T_{02}} \\ 
c_{I_{01}} \\ 
c_{T_{12}} \\ 
c_{T_{10}} \\ 
c_{I_{11}} \\ 
c_{T_{20}} \\ 
c_{T_{21}} \\ 
c_{I_{21}} \\ 
c_{C_0}    \\
c_{C_1}    \\
c_{C_2}    \\
c_{I_{02}} \\
c_{I_{12}} \\
c_{I_{22}} \\
c_S
\end{array}
}
\right]
\end{align}

\begin{align}
\nonumber {\bf R} &=
\left[
\scalemath{0.4}{
\begin{array}{*{20}{c}}
\frac{x_{V_1}-x_{V_0}}{3} & \frac{y_{V_1}-y_{V_0}}{3} & 0 & 0 & 0 & 0 & 0 & 0 & 0 \\ 
\frac{x_{V_2}-x_{V_0}}{3} & \frac{y_{V_2}-y_{V_0}}{3} & 0 & 0 & 0 & 0 & 0 & 0 & 0 \\ 
\frac{x_{V_1}+x_{V_2}-2x_{V_0}}{9} & \frac{y_{V_1}+y_{V_2}-2y_{V_0}}{9} & 0 & 0 & 0 & 0 & 0 & 0 & 0 \\ 
0 & 0 & \frac{x_{V_2}-x_{V_1}}{3} & \frac{y_{V_2}-y_{V_1}}{3} & 0 & 0 & 0 & 0 & 0 \\  
0 & 0 & \frac{x_{V_0}-x_{V_1}}{3} & \frac{y_{V_0}-y_{V_1}}{3} & 0 & 0 & 0 & 0 & 0 \\ 
0 & 0 & \frac{x_{V_0}+x_{V_2}-2x_{V_1}}{9} & \frac{y_{V_0}+y_{V_2}-2y_{V_1}}{9} & 0 & 0 & 0 & 0 & 0 \\ 
0 & 0 & 0 & 0 & \frac{x_{V_0}-x_{V_2}}{3} & \frac{y_{V_0}-y_{V_2}}{3} & 0 & 0 & 0 \\ 
0 & 0 & 0 & 0 & \frac{x_{V_1}-x_{V_2}}{3} & \frac{y_{V_1}-y_{V_2}}{3} & 0 & 0 & 0 \\ 
0 & 0 & 0 & 0 & \frac{x_{V_0}+x_{V_1}-2x_{V_2}}{9} & \frac{y_{V_0}+y_{V_1}-2y_{V_2}}{9} & 0 & 0 & 0 \\ 
0 & 0 & \frac{\theta_{21}(x_{V_2}-x_{V_1})}{3} & \frac{\theta_{21}(y_{V_2}-y_{V_1})}{3} & \frac{\theta_{12}(x_{V_1}-x_{V_2})}{3} & \frac{\theta_{12}(y_{V_1}-y_{V_2})}{3} & \eta_0 & 0 & 0 \\ 
\frac{\theta_{20}(x_{V_2}-x_{V_0})}{3} & \frac{\theta_{20}(y_{V_2}-y_{V_0})}{3} & 0 & 0 & \frac{\theta_{02}(x_{V_0}-x_{V_2})}{3} & \frac{\theta_{02}(y_{V_0}-y_{V_2})}{3} & 0 & \eta_1 & 0 \\ 
\frac{\theta_{01}(x_{V_1}-x_{V_0})}{3} & \frac{\theta_{01}(y_{V_1}-y_{V_0})}{3} & \frac{\theta_{10}(x_{V_0}-x_{V_1})}{3} & \frac{\theta_{10}(y_{V_0}-y_{V_1})}{3} & 0 & 0 & 0 & 0 & \eta_2 \\ 
\frac{(3\theta_{01}+1)x_{V_1}+(3\theta_{20}+1)x_{V_2}-(3\theta_{20}+3\theta_{01}+2)x_{V_0}}{27} & \frac{(3\theta_{01}+1)y_{V_1}+(3\theta_{20}+1)y_{V_2}-(3\theta_{20}+3\theta_{01}+2)y_{V_0}}{27} & \frac{\theta_{10}(x_{V_0}-x_{V_1})}{9} & \frac{\theta_{10}(y_{V_0}-y_{V_1})}{9} & \frac{\theta_{02}(x_{V_0}-x_{V_2})}{9} & \frac{\theta_{02}(y_{V_0}-y_{V_2})}{9} & 0 & \frac{\eta_1}{3} & \frac{\eta_2}{3} \\ 
\frac{\theta_{01}(x_{V_1}-x_{V_0})}{9} & \frac{\theta_{01}(y_{V_1}-y_{V_0})}{9} & \frac{(3\theta_{10}+1)x_{V_0}+(3\theta_{21}+1)x_{V_2}-(3\theta_{10}+3\theta_{21}+2)x_{V_1}}{27} & \frac{(3\theta_{10}+1)y_{V_0}+(3\theta_{21}+1)y_{V_2}-(3\theta_{10}+3\theta_{21}+2)y_{V_1}}{27} & \frac{\theta_{12}(x_{V_1}-x_{V_2})}{9} & \frac{\theta_{12}(y_{V_1}-y_{V_2})}{9} & \frac{\eta_0}{3} & 0 & \frac{\eta_2}{3} \\
\frac{\theta_{20}(x_{V_2}-x_{V_0})}{9} & \frac{\theta_{20}(y_{V_2}-y_{V_0})}{9} & \frac{\theta_{21}(x_{V_2}-x_{V_1})}{9} & \frac{\theta_{21}(y_{V_2}-y_{V_1})}{9} & \frac{(3\theta_{02}+1)x_{V_0}+(3\theta_{12}+1)x_{V_1}-(3\theta_{02}+3\theta_{12}+2)x_{V_2}}{27} & \frac{(3\theta_{02}+1)y_{V_0}+(3\theta_{12}+1)y_{V_1}-(3\theta_{02}+3\theta_{12}+2)y_{V_2}}{27} & \frac{\eta_0}{3} & \frac{\eta_1}{3} & 0 \\
\frac{(6\theta_{02}+1)x_{V_0}+(6\theta_{12}+1)x_{V_1}-(6\theta_{02}+6\theta_{12}+2)x_{V_2}}{27} & \frac{(6\theta_{02}+1)y_{V_0}+(6\theta_{12}+1)y_{V_1}-(6\theta_{02}+6\theta_{12}+2)y_{V_2}}{27} & \frac{(6\theta_{10}+1)x_{V_0}+(6\theta_{21}+1)x_{V_2}-(6\theta_{10}+6\theta_{21}+2)x_{V_1}}{27} & \frac{(6\theta_{10}+1)y_{V_0}+(6\theta_{21}+1)y_{V_2}-(6\theta_{10}+6\theta_{21}+2)y_{V_1}}{27} & \frac{(6\theta_{02}+1)x_{V_0}+(6\theta_{12}+1)x_{V_1}-(6\theta_{02}+6\theta_{12}+2)x_{V_2}}{27} & \frac{(6\theta_{02}+1)y_{V_0}+(6\theta_{12}+1)y_{V_1}-(6\theta_{02}+6\theta_{12}+2)y_{V_2}}{27} & \frac{2\eta_0}{3} & \frac{2\eta_1}{3} & \frac{2\eta_2}{3} 
\end{array}
}
\right]
\end{align}

\begin{align}
\nonumber {\bf d} &=
\left[
\scalemath{0.9}{
\begin{array}{*{20}{c}}
d_{V_0}^x \\
d_{V_0}^y \\
d_{V_1}^x \\
d_{V_1}^y \\
d_{V_2}^x \\
d_{V_2}^y \\
d_{E_0}^e \\
d_{E_1}^e \\
d_{E_2}^e
\end{array}
}
\right]
\end{align}

\begin{align}
\nonumber {\bf f} &=
\left[
\scalemath{0.9}{
\begin{array}{*{20}{c}}
z_{V_0} \\
z_{V_0} \\
z_{V_0} \\
z_{V_1} \\
z_{V_1} \\
z_{V_1} \\
z_{V_2} \\
z_{V_2} \\
z_{V_2} \\
\theta_{21}z_{V_1}+\theta_{12}z_{V_2} \\
\theta_{02}z_{V_2}+\theta_{20}z_{V_0} \\
\theta_{10}z_{V_0}+\theta_{01}z_{V_1} \\
\frac{(\theta_{20}+\theta_{10}+1)z_{V_0}+\theta_{21}z_{V_1}+\theta_{02}z_{V_2}}{3} \\
\frac{\theta_{10}z_{V_0}+(\theta_{21}+\theta_{01}+1)z_{V_1}+\theta_{12}z_{V_2}}{3} \\
\frac{\theta_{20}z_{V_0}+\theta_{01}z_{V_1}+(\theta_{12}+\theta_{02}+1)z_{V_2}}{3} \\
\frac{(2\theta_{20}+2\theta_{10}+1)z_{V_0}+(2\theta_{21}+2\theta_{01}+1)z_{V_1}+(2\theta_{12}+2\theta_{02}+1)z_{V_2}}{3}
\end{array}
}
\right]
\end{align}

\section{Details of Equality Constraint}
Summarizing (5) as matrix form, we obtain
\begin{align}
\nonumber {\bf G}{\bf c} \leq {\bf h},
\end{align}
where
\begin{align}
\nonumber {\bf G} = &\left[
\scalemath{0.6}{
\begin{array}{*{20}{c}}
y_{V_0}-y_{V_2} & y_{V_1}-y_{V_0} & 0 & 0 & 0 & 0 & 0 & 0 & 0 & 0 & 0 & 0 & 0 & 0 & 0 & 0 \\ 
0 & 0 & 0 & y_{V_1}-y_{V_0} & y_{V_2}-y_{V_1} & 0 & 0 & 0 & 0 & 0 & 0 & 0 & 0 & 0 & 0 & 0 \\ 
0 & 0 & 0 & 0 & 0 & 0 & y_{V_2}-y_{V_1} & y_{V_0}-y_{V_2} & 0 & 0 & 0 & 0 & 0 & 0 & 0 & 0 \\ 
0 & 0 & y_{V_2}-y_{V_1} & 0 & 0 & 0 & 0 & 0 & 0 & 0 & y_{V_1}-y_{V_0} & y_{V_0}-y_{V_2} & 0 & 0 & 0 & 0 \\ 
0 & 0 & 0 & 0 & 0 & y_{V_0}-y_{V_2} & 0 & 0 & 0 & y_{V_1}-y_{V_0} & 0 & y_{V_2}-y_{V_1} & 0 & 0 & 0 & 0 \\ 
0 & 0 & 0 & 0 & 0 & 0 & 0 & 0 & y_{V_1}-y_{V_0} & y_{V_0}-y_{V_2} & y_{V_2}-y_{V_1} & 0 & 0 & 0 & 0 & 0 \\ 
0 & 0 & 0 & 0 & 0 & 0 & 0 & 0 & 0 & 0 & 0 & 0 & y_{V_2}-y_{V_1} & y_{V_0}-y_{V_2} & y_{V_1}-y_{V_0} & 0 \\ 
y_S-y_{V_1} & 0 & 0 & 0 & y_{V_0}-y_S & 0 & 0 & 0 & 0 & 0 & 0 & y_{V_1}-y_{V_0} & 0 & 0 & 0 & 0 \\ 
0 & 0 & 0 & y_S-y_{V_2} & 0 & 0 & 0 & y_{V_1}-y_S & 0 & y_{V_2}-y_{V_1} & 0 & 0 & 0 & 0 & 0 & 0 \\ 
0 & y_{V_2}-y_S & 0 & 0 & 0 & 0 & y_S-y_{V_0} & 0 & 0 & 0 & y_{V_0}-y_{V_2} & 0 & 0 & 0 & 0 & 0
\end{array}
}
\right],
\end{align}
and 
\begin{align}
\nonumber {\bf h} = \left[
\begin{array}{*{20}{c}}
(y_{V_1}-y_{V_2})z_{V_0} \\ 
(y_{V_2}-y_{V_0})z_{V_1} \\ 
(y_{V_0}-y_{V_1})z_{V_2} \\ 
0 \\ 
0 \\ 
0 \\ 
0 \\ 
0 \\ 
0 \\ 
0
\end{array}
\right].
\end{align}

\section{Details of Loss Function}
Expanding (14), we obtain
\begin{align}
\nonumber U=&\left[
\scalemath{0.7}{
\begin{array}{*{20}{c}}
\frac{18}{||E_2||^3} & 0 & 0 & 0 & \frac{-9}{||E_2||^3} & 0 & 0 & 0 & 0 & 0 & 0 & 0 & 0 & 0 & 0 & 0 \\ 
0 & \frac{18}{||E_1||^3} & 0 & 0 & 0 & 0 & \frac{-9}{||E_1||^3} & 0 & 0 & 0 & 0 & 0 & 0 & 0 & 0 & 0 \\ 
0 & 0 & \frac{36}{||\hat{E}_0||^3} & 0 & 0 & 0 & 0 & 0 & 0 & 0 & 0 & 0 & \frac{-18}{||\hat{E}_0||^3} & 0 & 0 & 0 \\ 
0 & 0 & 0 & \frac{18}{||E_0||^3} & 0 & 0 & 0 & \frac{-9}{||E_0||^3} & 0 & 0 & 0 & 0 & 0 & 0 & 0 & 0 \\ 
\frac{-9}{||E_2||^3} & 0 & 0 & 0 & 0 & 0 & 0 & 0 & 0 & 0 & 0 & 0 & 0 & 0 & 0 & 0 \\ 
0 & 0 & 0 & 0 & 0 & \frac{36}{||\hat{E}_1||^3} & 0 & 0 & 0 & 0 & 0 & 0 & 0 & \frac{-18}{||\hat{E}_1||^3} & 0 & 0 \\ 
0 & \frac{-9}{||E_1||^3} & 0 & 0 & 0 & 0 & \frac{18}{||E_1||^3} & 0 & 0 & 0 & 0 & 0 & 0 & 0 & 0 & 0 \\ 
0 & 0 & 0 & \frac{-9}{||E_0||^3} & 0 & 0 & 0 & \frac{18}{||E_0||^3} & 0 & 0 & 0 & 0 & 0 & 0 & 0 & 0 \\ 
0 & 0 & 0 & 0 & 0 & 0 & 0 & 0 & \frac{36}{||\hat{E}_2||^3} & 0 & 0 & 0 & 0 & 0 & \frac{-18}{||\hat{E}_2||^3} & 0 \\ 
0 & 0 & 0 & 0 & 0 & 0 & 0 & 0 & 0 & 0 & 0 & 0 & 0 & 0 & 0 & 0 \\
0 & 0 & 0 & 0 & 0 & 0 & 0 & 0 & 0 & 0 & 0 & 0 & 0 & 0 & 0 & 0 \\
0 & 0 & 0 & 0 & 0 & 0 & 0 & 0 & 0 & 0 & 0 & 0 & 0 & 0 & 0 & 0 \\
0 & 0 & \frac{-18}{||\hat{E}_0||^3} & 0 & 0 & 0 & 0 & 0 & 0 & 0 & 0 & 0 & \frac{36}{||\hat{E}_0||^3} & 0 & 0 & \frac{-18}{||\hat{E}_0||^3} \\
0 & 0 & 0 & 0 & 0 & \frac{-18}{||\hat{E}_1||^3} & 0 & 0 & 0 & 0 & 0 & 0 & 0 & \frac{36}{||\hat{E}_1||^3} & 0 & \frac{-18}{||\hat{E}_1||^3} \\
0 & 0 & 0 & 0 & 0 & 0 & 0 & 0 & \frac{-18}{||\hat{E}_2||^3} & 0 & 0 & 0 & 0 & 0 & \frac{36}{||\hat{E}_2||^3} & \frac{-18}{||\hat{E}_2||^3} \\
0 & 0 & 0 & 0 & 0 & 0 & 0 & 0 & 0 & 0 & 0 & 0 & \frac{-18}{||\hat{E}_0||^3} & \frac{-18}{||\hat{E}_1||^3} & \frac{-18}{||\hat{E}_2||^3} & \frac{12}{||\hat{E}_0||^3}+\frac{12}{||\hat{E}_1||^3}+\frac{12}{||\hat{E}_2||^3}
\end{array}
}
\right],
\end{align}
and
\begin{align}
\nonumber {\bf w}_{V_0V_1V_2} = &\left[
\scalemath{0.7}{
\begin{array}{*{20}{c}}
\frac{-18z_{V_0}}{||E_2||^3} \\
\frac{-18z_{V_0}}{||E_1||^3} \\
\frac{-36z_{V_0}}{||\hat{E}_0||^3} \\
\frac{-18z_{V_1}}{||E_0||^3} \\
\frac{-18z_{V_1}}{||E_2||^3} \\
\frac{-36z_{V_1}}{||\hat{E}_1||^3} \\
\frac{-18z_{V_2}}{||E_1||^3} \\
\frac{-18z_{V_2}}{||E_0||^3} \\
\frac{-36z_{V_2}}{||\hat{E}_2||^3} \\
0 \\
0 \\
0 \\
0 \\
0 \\
0 \\
\frac{12z_{V_0}}{||\hat{E}_0||^3}+\frac{12z_{V_1}}{||\hat{E}_1||^3}+\frac{12z_{V_2}}{||\hat{E}_2||^3}
\end{array}
}
\right].
\end{align}

\section{Visual Illustration}
\begin{figure*}[ht]
    \centering
    \captionsetup{justification=centering}
    \subfloat[CT 20 iterations]{\includegraphics[width=0.33\textwidth]{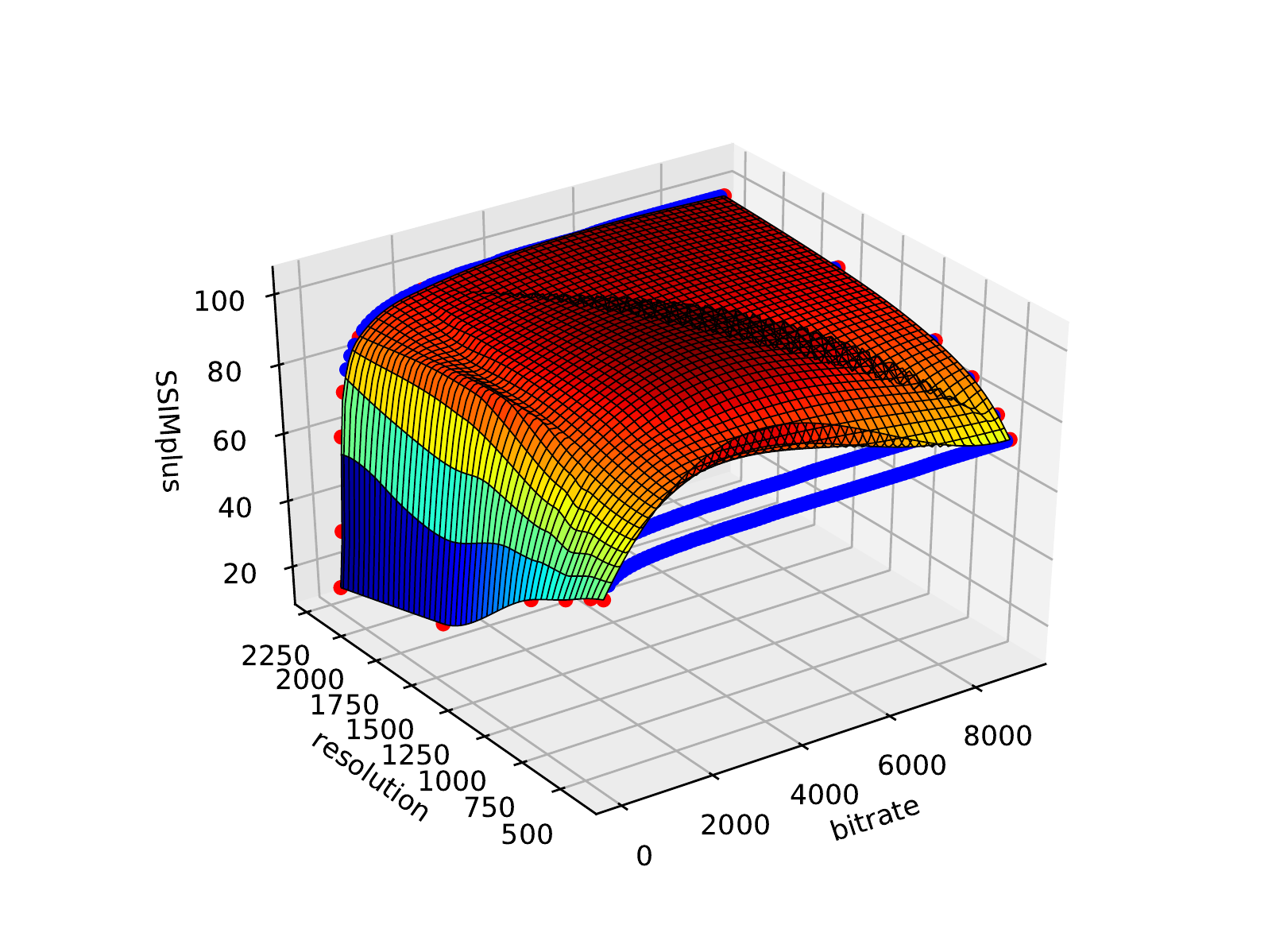}}\hskip.2em
    \subfloat[CT 30 iterations]{\includegraphics[width=0.33\textwidth]{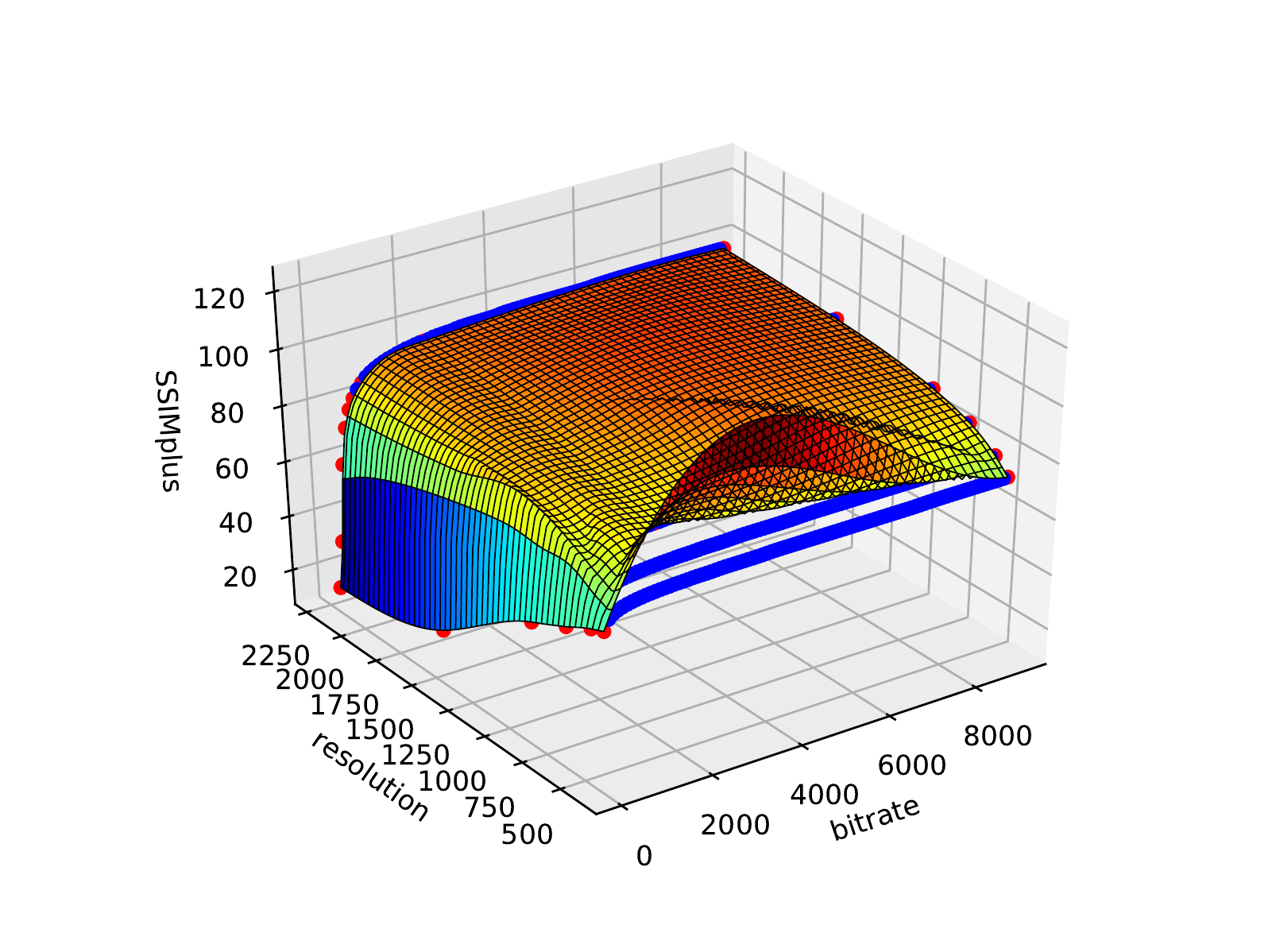}}\hskip.2em
    \subfloat[CT 40 iterations]{\includegraphics[width=0.33\textwidth]{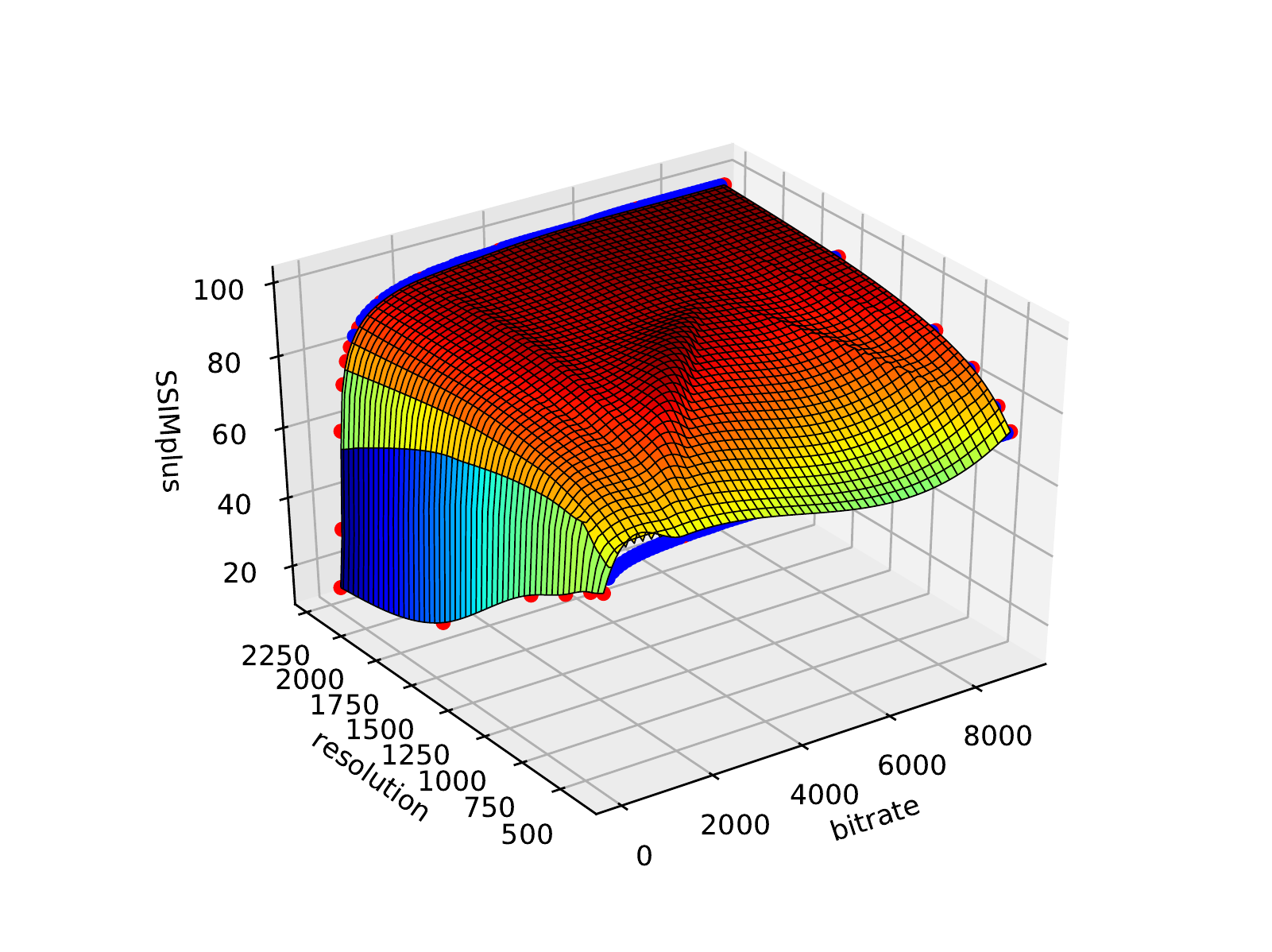}}

    \subfloat[RAMCT 20 iterations]{\includegraphics[width=0.33\textwidth]{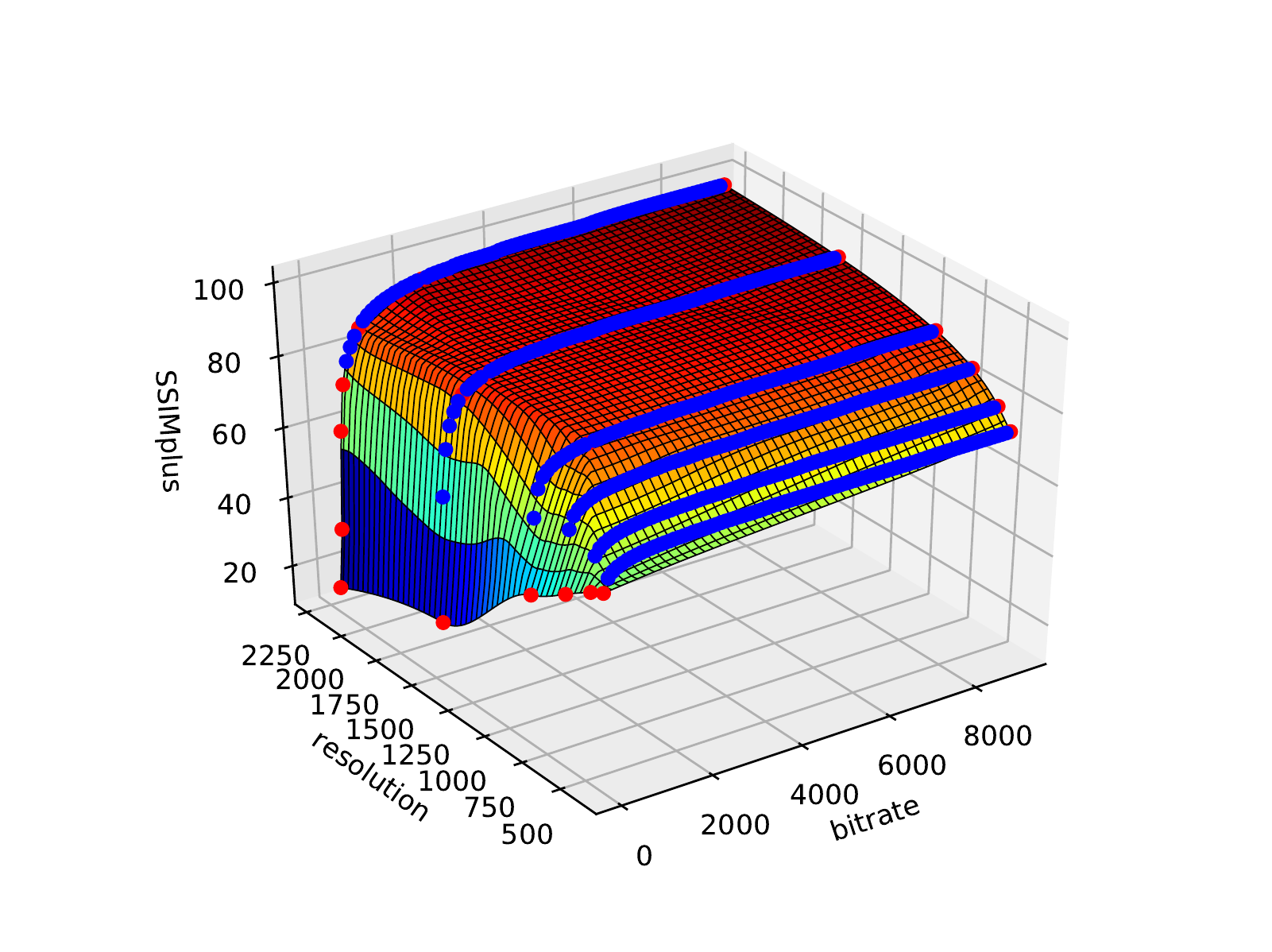}}\hskip.2em
    \subfloat[RAMCT 30 iterations]{\includegraphics[width=0.33\textwidth]{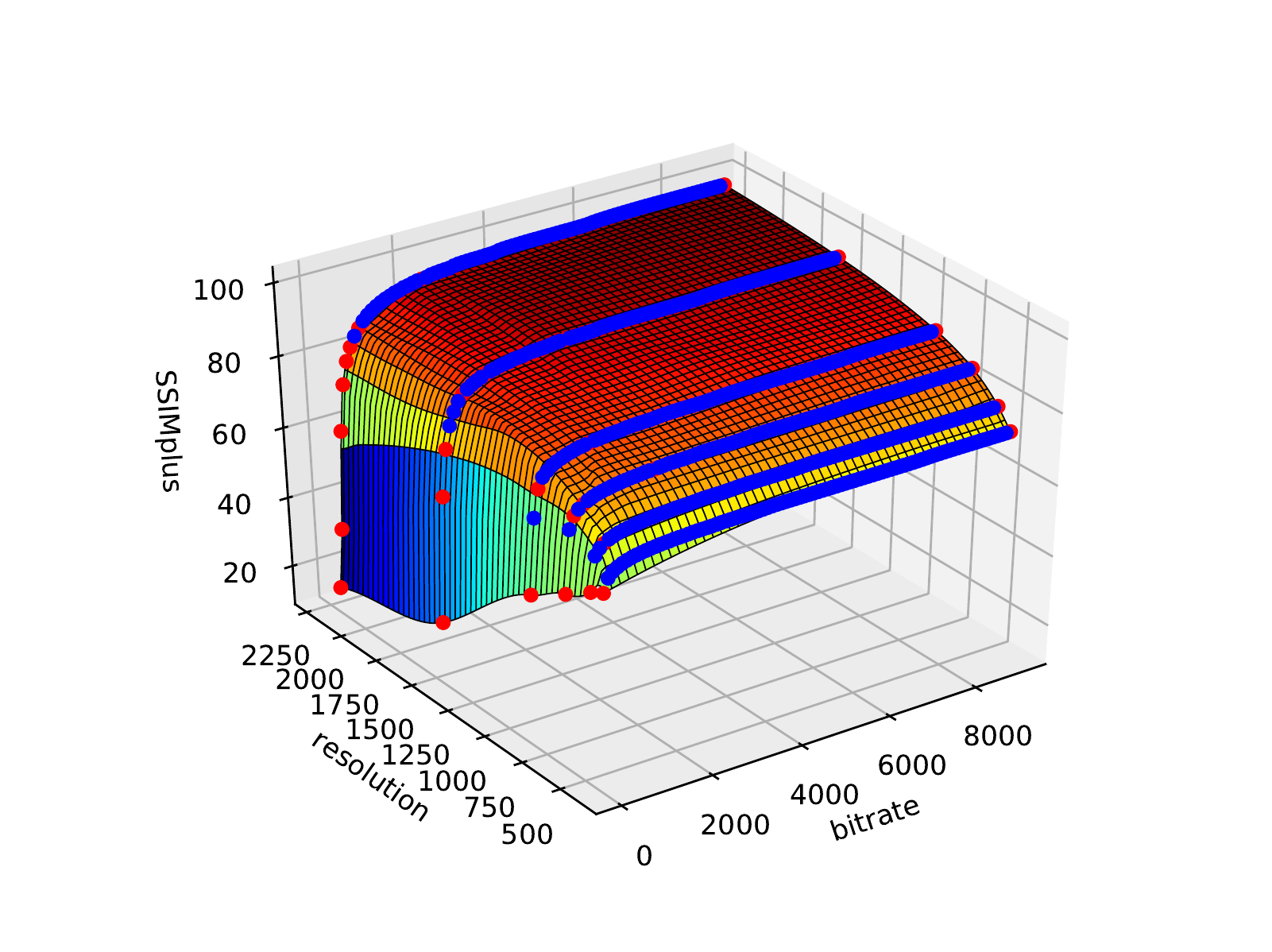}}\hskip.2em
    \subfloat[RAMCT 40 iterations]{\includegraphics[width=0.33\textwidth]{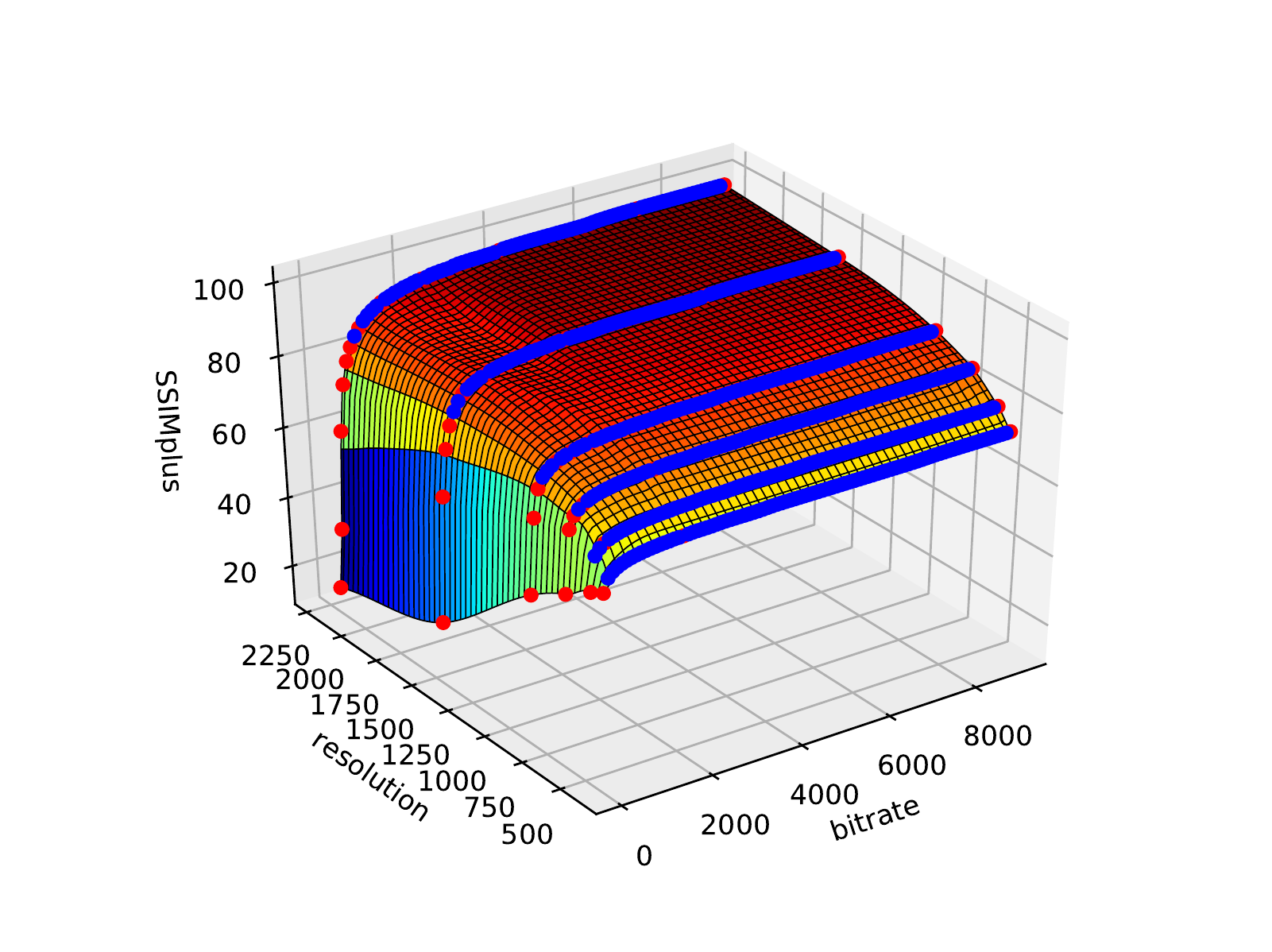}}
    \caption{CT and RAMCT performance given 20, 30, and 40 training samples selected by the proposed uncertainty sampling method. Red and blue points represent the training and testing ground truth quality scores, respectively.}\label{fig:performance}
\end{figure*}

\section{Proof of Necessary Conditions of Axial Monotonicity}
Since the interpolated surface is a continuous piece-wise cubic function, it being monotonic everywhere is equivalent to every cubic function in its own triangle being monotonic. Denote $(i, j, k)$ a cyclic permutation of $(0,1,2)$. Consider a microtriangle $\Delta_{V_iV_jS}$. The B\'ezier surface $z(\alpha, \beta, \gamma)$ over the microtriangle can be formulated as (1), where the barycentric coordinates $(\alpha, \beta, \gamma)$ are definded as in (2). The chain rule of derivatives shows that 
\begin{align} \tag{A.1}\label{eq:z_to_x}
\frac{\partial z}{\partial x} =& \frac{\partial z}{\partial \alpha} \frac{\partial \alpha}{\partial x} + \frac{\partial z}{\partial \beta} \frac{\partial \beta}{\partial x} + \frac{\partial z}{\partial\gamma} \frac{\partial \gamma}{\partial x}.
\end{align}

From (1), we have
\begin{subequations} \label{eqs:z_to_bary}
  \begin{align} 
    \frac{\partial z}{\partial \alpha} = & 3c_{V_i}\alpha^2 + 6c_{T_{ij}}\alpha\beta + 6c_{I_{i1}}\alpha\gamma + 3c_{T_{ji}}\beta^2 + 3c_{I_{i2}}\gamma^2 + 6c_{C_k}\beta\gamma \tag{A.2a} \\
    \frac{\partial z}{\partial \beta} = & 3c_{T_{ij}}\alpha^2 + 6c_{T_{ji}}\alpha\beta + 6c_{C_k}\alpha\gamma + 3c_{V_j}\beta^2 + 3c_{I_{j2}}\gamma^2 + 6c_{I_{j1}}\beta\gamma \tag{A.2b} \\
    \frac{\partial z}{\partial \gamma} = & 3c_{I_{i1}}\alpha^2 + 6c_{C_{k}}\alpha\beta + 6c_{I_{i2}}\alpha\gamma + 3c_{I_{j1}}\beta^2 + 3c_{S}\gamma^2 + 6c_{I_{j2}}\beta\gamma, \tag{A.2c}
  \end{align}
\end{subequations}
and from (2),
\begin{align}
    \frac{\partial \alpha}{\partial x} = \frac{y_{V_j} - y_S}{2A}, \frac{\partial \beta}{\partial x} = \frac{y_{S} - y_{V_i}}{2A}, \frac{\partial \gamma}{\partial x} = \frac{y_{V_i} - y_{V_j}}{2A} \tag{A.3}\label{eqs:bary_to_x}
  \end{align}
where $A$ denotes the area of $\Delta_{V_iV_jS}$. Substitute (A.2) and (\ref{eqs:bary_to_x}) into (\ref{eq:z_to_x}). After some rearrangements, we reach
\begin{align}
\frac{\partial z}{\partial x} = & \frac{3}{2A}[c_{V_0}(y_{V_j} - y_S) + c_{T_{ij}}(y_{S} - y_{V_i}) + c_{I_{i1}}(y_{V_i} - y_{V_j})] \alpha^2 
+ \frac{3}{2A}[c_{T_{ji}}(y_{V_j} - y_S) + c_{V_{j}}(y_{S} - y_{V_i}) + c_{I_{j1}}(y_{V_i} - y_{V_j})] \beta^2 \nonumber \\
+ & \frac{3}{2A}[c_{I_{i2}}(y_{V_j} - y_S) + c_{I_{j2}}(y_{S} - y_{V_i}) + c_{S}(y_{V_i} - y_{V_j})] \gamma^2 
+ \frac{6}{2A}[c_{T_{ij}}(y_{V_j} - y_S) + c_{T_{ji}}(y_{S} - y_{V_i}) + c_{C_{k}}(y_{V_i} - y_{V_j})] \alpha\beta \nonumber \\
+ & \frac{6}{2A}[c_{I_{i1}}(y_{V_j} - y_S) + c_{C_{k}}(y_{S} - y_{V_i}) + c_{I_{i2}}(y_{V_i} - y_{V_j})] \alpha\gamma 
+ \frac{6}{2A}[c_{C_{k}}(y_{V_j} - y_S) + c_{I_{j1}}(y_{S} - y_{V_i}) + c_{I_{j2}}(y_{V_i} - y_{V_j})] \beta\gamma. \tag{A.4}\label{eq:z_to_x_expand}
\end{align}

$z$ is $x$-axial monotonic if and only if (\ref{eq:z_to_x_expand}) is nonnegative within $\Delta{V_iV_jS}$. Note that $\alpha, \beta, \gamma$ are all nonnegative inside or on the edge of $\Delta{V_iV_jS}$. One sufficient condition for (\ref{eq:z_to_x_expand}) to be nonnegative is its all coefficients are nonnegative [Han Lu et al.], as indicated in (5). To show that (5a) and (5c) (i.e. (14a) and (14c)) are necessary conditions for $\frac{\partial z}{\partial x}$ to be nonnegative, we simply check the value of $\frac{\partial z}{\partial x}$ at three vertices of $\Delta_{V_iV_jS}$, where $(\alpha, \beta, \gamma)$ equals $(1,0,0)$, $(0,1,0)$, and $(0,0,1)$, respectively. Substituting the three barycentric coordinates back into (\ref{eq:z_to_x_expand}), we obtain three inequalities as the necessary condition in $\Delta{V_iV_jS}$
\begin{subequations}\label{eqs:necess_01S}
  \begin{align}
    c_{V_i}(y_{V_j} - y_S) + c_{T_{ij}}(y_{S} - y_{V_i}) + c_{I_{i1}}(y_{V_i} - y_{V_j}) \geq & 0 \tag{A.5a}\label{eq:necess_ijS_i} \\
    c_{T_{ji}}(y_{V_j} - y_S) + c_{V_{j}}(y_{S} - y_{V_i}) + c_{I_{j1}}(y_{V_i} - y_{V_j}) \geq & 0 \tag{A.5b}\\
    c_{I_{i2}}(y_{V_j} - y_S) + c_{I_{j2}}(y_{S} - y_{V_i}) + c_{S}(y_{V_i} - y_{V_j}) \geq & 0. \tag{A.5c}\label{eq:necess_ijS_S} 
  \end{align}
\end{subequations}

The inner-triangle $C^1$ continuity implies that points $V_i, T_{ij}, I_{i1}, T_{ik}$ are coplanar, \emph{i.e.} $c_{I_{i1}} = (c_{V_i} + c_{T_{ij}} + c_{T_ik})/3$. Substitute this equation and $y_S = (y_{V_i} + y_{V_j} + y_{V_k})/3$ into (\ref{eq:necess_ijS_i}). Some rearrangements will yield 
\begin{align}
c_{V_i}(y_{V_k} - y_{V_j}) + c_{T_{ij}}(y_{V_i} - y_{V_k}) + c_{T_{ik}}(y_{V_j} - y_{V_i}) & \leq 0, \tag{A.6}
\end{align}
which is exactly (5a). Further, the summation of (\ref{eq:necess_ijS_S}) over all possible $(i,j,k)$ gives (8c). Now we have proved (5a) and (5c) in Section 3.3 are necessary for the interplant to be axial monotonic in $\Delta{V_0V_1V_2}$.

\section{Details of Per-Title Optimization}
According to~\cite{zhang2013qoe}, the optimality condition of the constrained optimization problem in  is achieved when the following conditions hold. First, if the quality at the lowest bitrate is higher than the target quality $C_i$, the $i$-th optimal representation is $(x_L, y^*)$, where $x_L$ is the lowest bitrate and $y^* = argmax_{y} f(x_L, y)$. Second, when the target quality cannot be obtained by the lowest bitrate, the optimality condition is fulfilled when the inequality constraints are active (\textit{i.e.}, the equation condition is achieved). Consequently, the optimization problem can be transformed into finding the bitrate $x_i$ at each resolution such that $C_i = f(x_i, y)$, $\forall y$. We apply the Dichotomous-based search method~\cite{zhang2013qoe} to compute the optimal bitrate at {320$\times$240, 384$\times$288, 512$\times$384, 640$\times$360, 720$\times$480, 960$\times$540, 1280$\times$720, 1600$\times$900, 1920$\times$1080}~\cite{screensiz2018size}. The final bitrate is then obtained by taking the minimum value across all resolutions.

\section{Details of Generalized Bj{\o}ntegaard-Delta Models}
\subsection{Generalized Bj{\o}ntegaard-Delta Quality} We apply the $trapz$ function in Numpy~\cite{numpy2005} to compute the numerical integration in (23a) at each resolution from 500 kbps to 4000 kbps with 100 steps. The reason to select the bitrate range is that theoretically Bj{\o}ntegaard-Delta models should be computed based on actual bitrate instead of target bitrate. However, some encoders are not able to precisely control the encoding bitrate at very low or high bitrate ranges. We find the actual bitrate variability of encoders do not exceed 10\% of target bitrate within the this range, suggesting the error introduced by bitrate control is insignificant~\cite{applea}. The bitrate integration is computed at {320$\times$240, 384$\times$288, 512$\times$384, 640$\times$360, 720$\times$480, 960$\times$540, 1280$\times$720, 1600$\times$900, 1920$\times$1080}~\cite{screensiz2018size}, and integrated over spatial resolution to obtain the final score.

\begin{figure}
  \centering
  \includegraphics[width=0.5\textwidth]{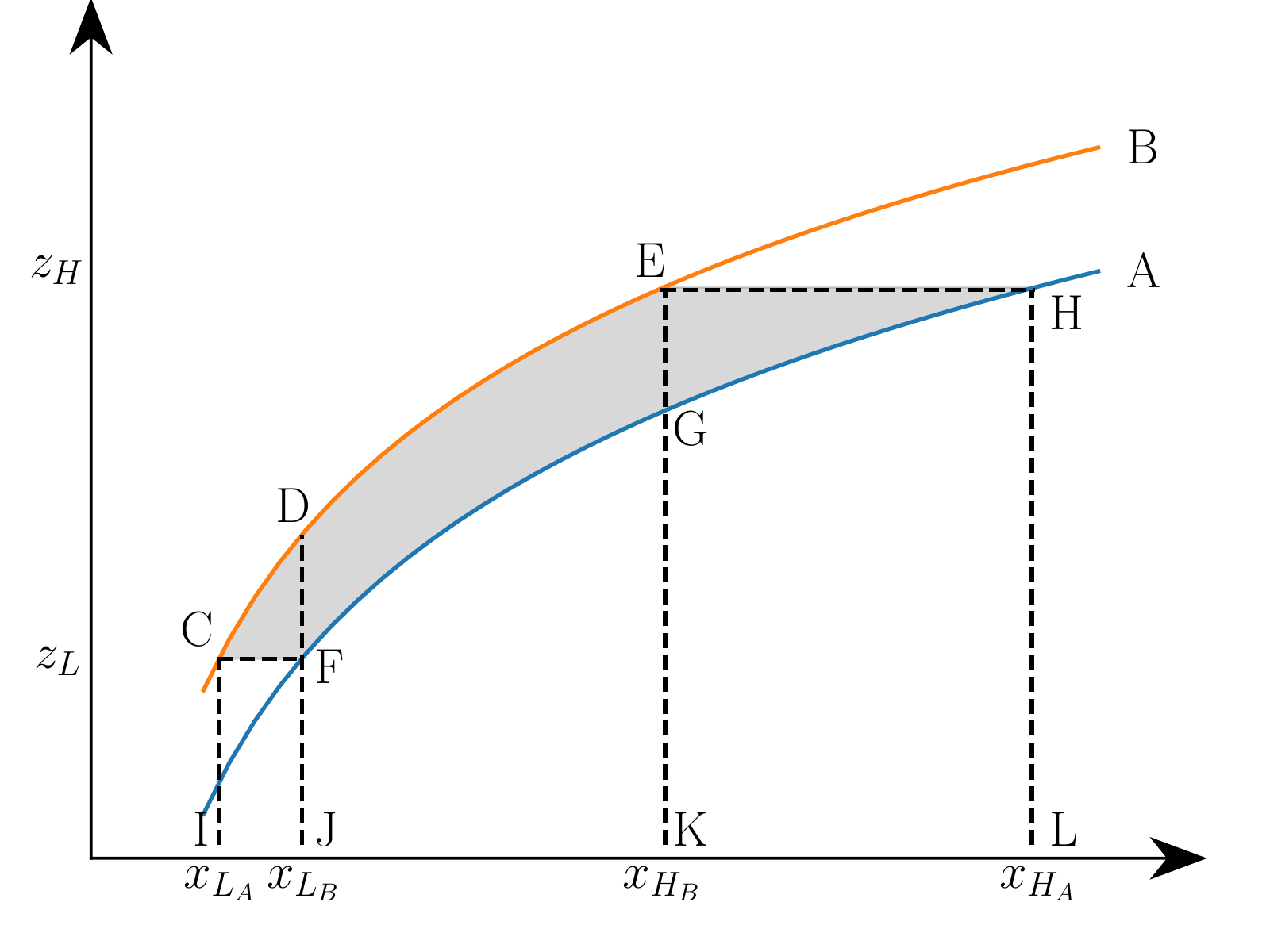}
  \caption{Computation procedure of generalized Bj{\o}ntegaard-Delta Rate at one spatial resolution. The shaded area represents the overall bitrate saving.}\label{fig:integral}
\end{figure}

\subsection{Generalized Bj{\o}ntegaard-Delta Rate} The computation procedure of generalized Bj{\o}ntegaard-Delta Rate at one resolution is illustrated in Fig.~\ref{fig:integral}. Specifically, the shaded area $S_{target}$ can be computed as
\begin{equation*}
S_{target} = AUC_{CDE} + S_{EHLK} - AUC_{FGH} - S_{CFJI},
\end{equation*}
where $S$ and $AUC$ stand for the area of rectangle and area under curve, respectively. The integrations require precise knowledge regarding to the bitrates $x_{L_A}$, $x_{L_B}$, $x_{H_A}$, $x_{H_A}$, and the effective quality interval $[z_L, z_H]$ at each resolution. At any resolution $y$, the integration interval is defined as
\begin{equation*}
\begin{aligned}
z_L&= max(z_A(x_L, y), z_B(x_L, y)) ,  \\
z_H&= min(z_A(x_H, y), z_B(x_H, y)) ,
\end{aligned}
\end{equation*}
in the original BD-Rate model. The bitrates ($x_{L_A}$, $x_{L_B}$, $x_{H_A}$, $x_{H_A}$) where the two rate-distortion functions achieve $z_L$ and $z_H$ are obtained with the Dichotomous-based search method~\cite{zhang2013qoe}. We apply the $trapz$ function in Numpy~\cite{numpy2005} to compute the numerical area under curves with step number of 100. In the end, we can integrate the overall bitrate saving over all spatial resolutions and devices to obtain the GBD-rate.